\documentclass{jfm}
\usepackage{tabularx}
\usepackage{graphicx}
\usepackage{upgreek}
\usepackage[square,sort,comma,numbers]{natbib}
\usepackage{pifont}
\usepackage{soul}
\usepackage{amsmath}
\usepackage{amssymb}
\usepackage{amsfonts}
\usepackage{color}
\usepackage{mathtools}
\usepackage{relsize}
\usepackage{wasysym}
\usepackage{fancyhdr}
\usepackage{setspace}
\usepackage[hidelinks=true, breaklinks=true]{hyperref}
\usepackage{url}
\include{Header}

\hypersetup{
    colorlinks = true,
    linkcolor  = blue,
    citecolor  = blue,      
    urlcolor   = blue,
}

\newcommand{\vol}{\mathop{\ooalign{\hfil$V$\hfil\cr\kern0.08em--\hfil\cr}}\nolimits}

\newcommand{\bs}[1]{\boldsymbol{#1}}

\newcommand{\mrm}[1]{\mathrm{#1}}

\newcommand{\imag}{\mathrm{i}}
\title[Evolution of asymmetries in trailing vortices]{Spatiotemporal evolution of asymmetries in turbulent trailing vortices}

\author[M. A. Khodkar]{M. A. Khodkar$^1$ \thanks{Email addresses for correspondence: ak152@rice.edu}}
\affiliation{$^1$Department of Mechanical Engineering, Rice University, Houston, TX 77005, USA}

\date{\today}

\pubyear{2022}
\volume{??}
\pagerange{???--???}
\date{?; revised ?; accepted ?. - To be entered by editorial office}

\begin{document}

\maketitle

\begin{abstract}

The outward propagation of asymmetries introduced to originally axisymmetric turbulent flows is investigated numerically and semi-analytically, where three-dimensional (3D) Batchelor vortices at high Reynolds numbers and with arbitrary swirl numbers are explored as test cases. It is well established that disturbances (asymmetries) added to a two-dimensional axisymmetric flow propagate radially outward, in order to re-axisymmetrize the vortex, but they cease to travel at a critical distance, known as the stagnation radius (\href{https://rmets.onlinelibrary.wiley.com/doi/10.1002/qj.49712353810}{Montgomery and Kallenbach, \textit{Q. J . R. Meteorol. Soc.}, vol. 123, 1997, pp. 435--465}). We utilize direct numerical simulations (DNS) and an inviscid model developed by linearizing the momentum and vorticity transport equations around the base (unperturbed) flow in helical coordinates to demonstrate that, in contrast with two-dimensional cases, 3D vortices enable the unbounded radial propagation of asymmetries. We further apply the Wenzel-Kramers-Brillouin (WKB) analysis to the linear model, which treats perturbations as compact wavepackets, to transform the partial differential equations of the linear model to a few ordinary differential equations. However it has been shown in the climate science community that the WKB approach ubiquitously predicts stagnation radii and heights for disturbances introduced to 3D cyclone-like vortices, here, we are able to identify a narrow range of parameters, for which the WKB analysis also supports an unrestrained, outward propagation for disturbances. Finally, the mechanisms governing the momentum transport at different times and locations, thereby promoting the outward advection of perturbations, are elucidated using the the linear rapid distortion theory (RDT) and numerical simulations. RDT is a powerful means for the study of vortex-turbulence interactions based on the separation of the flow into a steady, background field and a turbulent perturbation velocity field, which is initially homogeneous and isotropic. Since the DNS of the full Navier-Stokes equations rapidly stabilizes to a laminar, high-swirl-number configuration, the DNS of the linearized transport equations and the nonlinear governing equations without base-flow interactions are also carried out, in order to uncover the primary mechanisms for the growth and radial propagation of perturbations as well as the nonlinear processes causing growth arrest at fixed swirl numbers.        

\end{abstract}


%%%%%%%%%%%%%%%%%%%%%%%%%%%%%%%%%%%%%%%%%%%%%% Introduuction %%%%%%%%%%%%%%%%%%%%%%%%%%%%%%%%%%%%%%%%%%%%%%
%%%%%%%%%%%%%%%%%%%%%%%%%%%%%%%%%%%%%%%%%%%%%% Introduuction %%%%%%%%%%%%%%%%%%%%%%%%%%%%%%%%%%%%%%%%%%%%%%

\section{Introduction \label{section:Introduction}}         

The prevalence of large-scale vortices in engineering problems (e.g., trailing vortices in the wake of flyers and swirling jets in turbomachinery) and natural phenomena (e.g., cyclones and tornadoes) has made them the subject of extensive research in the past few decades \citep{Saffman1992, Wu2006, Khodkar2016, Khodkar2017}. In particular, helical instabilities arising in these flows have been vastly studied theoretically and numerically. \citet{Lessen1974} adopted a linear, normal-mode stabilty analysis to show that inviscid flows are destabilized when $q \lesssim 1.5$, where the swirl number $q$ indicates the relative magntitude of tangential (azimuthal) velocity to axial velocity in the base (unperturbed) flow. The authors also displayed that the strongest instability with the highest amplification level appear for $q \approx 0.85$. \citet{Mayer1992} extended the work of \citet{Lessen1974} to viscous flows and demonstrated that as Reynold number $Re$ grows, the critical swirl number below which helical instabilities emerge increases and nears that predicted by \citet{Lessen1974}. \citet{Fabre2004} further exhibited the existence of a class of viscous instabilities with a `centre-mode' behaviour for all swirl numbers, if $Re$ is sufficiently large. An exhaustive survey of various eigenmodes appearing in viscous Lamb-Oseen vortices (vortices for which base flow has no axial velocity or, equivalently, $q = \infty$) is presented in \citet{Fabre2006}. 

In addition, several numerical investigations have been conducted on swirling-jet instabilities using large-eddy simulations (LES) \citep{Ragab1995, Pantano2002} and direct numerical simulations (DNS) \citep{Qin1998, Duraisamy2006, Duraisamy2008}. All these studies agree that the formation of large eddies caused by helical instabilies redistribute the axial momentum deficit, thus producing velocity profiles that correspond to higher swirl numbers, thereby driving the flow to a laminar, stable state. \citet{Duraisamy2008} delves into the mechanisms governing the transport of angular and axial momenta at different stages of the flow, to reveal that the initial linear mechanism in which Reynolds normal stresses ($u_r^{\prime 2}$, $u_{\theta}^{\prime 2}$ and $u_z^{\prime 2}$) are predominant is saturated at later times and large distances from the core radius (the radial location of the maximum tangential velocity), and is replaced by a nonlinear mechanism wherein Reynolds shear stresses ($u_r' u'_{\theta}$ and $u_r' u_z'$) play the key role in the process of momentum transport. 

Apart from the modal instabilities reviewed above, which arise in the limited range of small swirl numbers, numerous studies have found alternative routes to perturbation growth and subsequent instabilities in swirling jets, which can occur universally. In fact, independent of the values of $Re$ and $q$, swirling jets support the algebraic growth of perturbations, also known as \textit{transient growth}, in short time spans, owing to the non-normality of the operators governing their linearized transport equations, even if normal-mode instability analysis signals the asymptotic decay of all eigenfunctions \citep{Miyazaki2000, Pradeep2006, Heaton2006, Heaton2007}. The algebraic growth means that the perturbations vary as $t^{\sigma}$ with $\sigma > 0$, rather than growing exponentiolly as seen in modal instabilities. \citet{Pradeep2006} demonstrated that transient growth in three-dimensional (3D) Lamb-Oseen vortices is promoted by a mechanism hinging on the change in the alignment of vortex filaments from the radial to the azimuthal direction and the stretching of these filaments by the mean strain field, which helps $\omega_{\theta}$ grow and, consequently, generates $u_r' u'_{\theta}$. The authors also revealed that optimal modes (modes with maximum amplification), localized outside but near the vortex core, are produced by a resonance mechanism, when the mode oscillation frequency equals the angular velocity of the base flow. In an ensuing study, \citet{Hussain2011} illustrated that nonlinar effects suppress both the amplitude and duration of transient growth by intensifying the outward self-advection of vortex filaments near the core, which in turn removes perturbations from the vicinity of the resonant radius and terminates the resonance-driven amplification. \citet{Antkowiak2007} identified an amplification mechanism for Lamb-Oseen vortices whereby the radial Coriolis forces caused by the simultaneous presence of the base flow's mean rotation and the azimuthal velocity perturbations enable the transformation of azimuthal velocity streaks into vortex rings (rolls). This scenario can further be applied to a turbulent background to shed light on the underlying physics of nonlinear phenomena such as vortex breakdown and the associated turbulent amplification. However the process discovered by \citet{Antkowiak2007} is in clear constrast with the `lift-up' mechanism of planar shear flows, in which cross-flow disturbances (rolls) give rise to elongated streamwise structures (streaks), the energy amplification in both flows scales as $Re^2$. Furthermore, \citet{Heaton2007} expanded the work of \citet{Pradeep2006} by investigating swirling jets with finite swirl numbers (Batchelor vortices) and showed that transient energy growth is enhanced as $q$ decreases, and becomes particularly substantial when helical instabilities form ($q \lesssim 1.5$). 

In this paper, we explore the spatiotemporal evolution of asymmetries, also referred to as disturbances or perturbations henceforth, introduced to a 3D Batchelor vortex with an arbitrary swirl number, which, in the absence of disturbances, is axisymmetric. We leverage several analytical and computational tools to give insight into the underlying physics of interactions between turbulent vortices and disturbances. The numerical simulations are conducted using the DNS solver of \citet{Duraisamy2008}. We also construct an inviscid, two-dimmensional (2D) model by linearizing the momentum and vorticity transport equations around the base flow (the unpertubed Batchelor vortex) in helical coordinates, whose results are shown to agree closely with the DNS observations, both revealing that, unlike for 2D flows, the radially outward propagation of disturbances in 3D trailing vortices is not halted at any certain radius. The Wenzel-Kramers-Brillouin (WKB) analysis is then applied to the linear model, in order to further simplify its partial differential equations to a set of more tractable ordinary differential equations (ODEs). The WKB analysis has long been used as a capable analytical framework in the climate science and fluid dynamics communities, e.g., for the study of vortex axisymmetrization in hurricans and cyclones \citep{Montgomery1997, Moller2000, Gao2016} and the stability analysis of swirling jets \citep{Billant2005, Billant2013}. Employing the WKB approach for both barotropic and baroclinic cyclone-like vortices has rendered critical surfaces beyond which disturbances cannot travel \citep{Moller2000, Gao2016}, but, in this paper, we show that under certain conditions for the base flow parameters and the perturbation wavenumbers, the WKB analysis also predicts an unbounded, radial propagation for disturbances. Furthermore, the primary mechanisms of momentum transport, through which disturbances are convected outward, are uncovered at different stages and regions of the flow via DNS and the linear rapid distortion theory (RDT). The latter is a skilled methodology for the estamition of vortex-turbulence interactions without requiring computationally expensive numerical simulations \citep{Miyazaki2000}. 

The paper is organized as follows. The problem setup and the numerical solver are described in \S \ref{section:Setup}. The derivation of the PDEs of the 2D linear model and their reduction to a set of ODEs by the WKB analysis are presented in \S \ref{section:Theory}. The results of the linear model and the WKB approach are discussed in \S \ref{section:Results}. The underlying mechanisms enabling the outward proapagation of disturbances are explored in \S \ref{section:RDT} and \S \ref{section:DNS} using the RDT analysis and the numerical simulations, respectively. The main findings and concluding remarks are summarized in \S \ref{section:Conclusions}.         
        
%%%%%%%%%%%%%%%%%%%%%%%%%%%%%%%%%%%%%%%%%%%%%% Problem Setup %%%%%%%%%%%%%%%%%%%%%%%%%%%%%%%%%%%%%%%%%%%%%%
%%%%%%%%%%%%%%%%%%%%%%%%%%%%%%%%%%%%%%%%%%%%%% Problem Setup %%%%%%%%%%%%%%%%%%%%%%%%%%%%%%%%%%%%%%%%%%%%%%

\section{Problem setup and numerical simulations \label{section:Setup}}

The axisymmetric Batchelor vortex, originally introduced by \citet{Batchelor1964}, has been widely used as a base flow in numerous experimental and computational works on trailing vortices and swirling jets. The velocity and vorticity profiles of an unperturbed Batchelor vortex at $t = 0$ and in nondimensional form are characterized by
\begin{eqnarray}
	u_r  &=&  0 \, , \quad u_{\theta} = \frac{1 - \exp(-r^2)}{r} \, , \quad u_z = \frac{1}{q} \exp(-r^2)  \, , \nonumber \\           \omega_r  &=&  0 \, , \quad \omega_{\theta} = \frac{2r}{q} \exp(-r^2) \, , \quad \ \ \omega_z = 2\exp(-r^2) \, , \label{Batchelor_base0}
\end{eqnarray} 
where $1/\sqrt{\alpha}$ times unit length (with $\alpha = 1.2564$) and the initial axial velocity at the origin when $q = 1$, namely, $u_0$, have been selected as the characteristic length and velocity scales, respectively. The choice of the Lamb's constant $\alpha$ ensures that the initial core radius ($r_{co}$) in dimensional form equals unity. Note that $r_{co}$ is used to denote the dimensionless core radius. Time is in turn nondimensionalized by $t_s = 1/(\sqrt{\alpha}u_0)$, which is by a factor of $2\pi \sqrt{\alpha}$ smaller than the eddy turnover time. We stress that all quantities and equations in this paper are nondimensional, unless otherwise stated. In the absence of any disturbances, the radial transport of momentum is solely governed by diffusion and, consequently, the temporal evolution of unperturbed velocity and vorticity fields are rendered by
\begin{eqnarray}
	u_r  &=&  0 \, , \quad u_{\theta} = \frac{1 - \exp\big[-(r/l)^2 \big]}{r} \, , \quad u_z = \frac{1}{qR^2} \exp\big[-(r/l)^2 \big]  \, , \nonumber \\           \omega_r  &=&  0 \, , \quad \omega_{\theta} = \frac{2r}{ql^4} \exp\big[-(r/l)^2 \big] \, , \quad \ \ \omega_z = \frac{2}{l^2}\exp\big[-(r/l)^2 \big] \, , \label{Batchelor_base}
\end{eqnarray}
where $l = 1 + 4t/(Re/2\pi)$. The Reynolds number $Re$ is defined as $\Gamma/\nu = 2\pi u_0/(\sqrt{\alpha} \nu)$, with $\Gamma$ and $\nu$ representing the kinematic viscosity, respectively. Eq.~(\ref{Batchelor_base}) implies that in the limit of inviscid flow the velocity and vorticity profiles of the Batchloer vortex become time-independent. Once asymmetries are introduced, however, rapidly rotating vortices expand, thereby advecting the perturbations outward \citep{Montgomery1997, Moller2000}. In this scenario, velocity and vorticity fields, which are substantially different from those descirbed by Eq.~(\ref{Batchelor_base}), can be decomposed into unperturbed (baseline) and perturbation parts
\begin{equation}
	 u_i = \overline{u}_i + u_i' \, , \quad \omega_i = \overline{\omega}_i + \omega_i' \, , \label{decompose} 
\end{equation}
indicated by overbar and prime, respectively. Here, $i$ can be replaced by $r$, $\theta$ and $z$. Note that the baseline values are the same as those in Eq.~(\ref{Batchelor_base}).   

The DNS solver used in this study employs a pseudospectral technique to solve the vorticity conservation equations, which in the vector form read
\begin{equation}
	\frac{\partial \bs{\omega}}{\partial t} + \bnabla \times (\bs{\omega} \times \bs{u}) = \frac{1}{Re} \nabla^2 \bs{\omega} \, . \label{DNS}
\end{equation}
Focusing on vorticity, instead of velocity, enables the efficient adoption of the pseudospectral approach, since the exponential decay of vorticity components, in contrast with the $1/r$ decrease in $u_{\theta}$, allows for the enforcement of periodic boundary conditions in a spatially compact fashion. Following the work of \citet{Rennich1997}, boundary conditions in the spanwise and vertical directions (corresponding to the $x$- and $y$-directions, respectively) are dealt with by decomposing the velocity vector $\bs{u}$ into an irrotational part, given by the potential function $\phi$, and a rotational part calculated from the mean axial circulation $A$, so that
\begin{equation}
	\bs{u} = \bnabla \phi + \frac{A}{2\pi r} \bs{e}_{\theta} + B \bs{e}_z \, , \label{DNS_bc}
\end{equation}
where $\bs{e}_{\theta}$ and $\bs{e}_{z}$ are the unit vectors in the tangential and axial directions, respectively, and $B$ guarantees that the freestream axial velocity is zero. In all simulations performed in this study, the size of the computational domain, the mesh size and the Reynolds number are selected as $L_x \times L_y \times L_z = (24 r_{co}/\sqrt{\alpha})^3$, $N_x \times N_y \times N_z = 384^3$ and $Re = 12 500$, respectively. More details on the present numerical model can be found in \citet{Duraisamy2006} and \citet{Duraisamy2008}. 

As discussed in the introduction, the DNS of unstable cases quickly laminarizes, while $q$ continually increases from its value at $t = 0$, prohibiting us from investigating the turbulent evolution of vortices at low swirl numbers for a sufficiently long time. In order to circumvent this problem, the numerical simulations of the linearized governing equations are conducted as well, wherein $\bnabla \times (\overline{\bs{\omega}} \times \overline{\bs{u}})$ and $\bnabla \times (\bs{\omega}' \times \bs{u}')$ terms produced by $\bnabla \times (\bs{\omega} \times \bs{u})$ have been neglected. This modified configuration, hereafter referred to as L-DNS, enables the study of the physical mechanisms responsible for the transient growth and outward advection of perturbations, as it retains the swirl number at its initially assigned value. The nonlinear mechanisms accounting for the cessation of growth and outward propgation of perturbations are explored via a numerical model, called N-DNS, which incorporates the interactions between velocity and vorticity perturbations (turbulence-turbulence interactions) by including the nonlinear terms in the form $\bnabla \times (\bs{\omega}' \times \bs{u}')$ appearing in the vorticity transport equation. Note, however, that N-DNS still ignores the base-flow interactions formulated as $\bnabla \times (\overline{\bs{\omega}} \times \overline{\bs{u}})$ and, as a consequence, does not capture the mean flow changes. This enables N-DNS to `freeze' the swirl number at its initial value, thus allowing for the investigation of nonlinear processes causing growth arrest at a given value of $q$ for a long enough time.

%%%%%%%%%%%%%%%%%%%%%%%%%%%%%%%%%%%%%%%%%%%%%% Theory (Linear Model and WKB) %%%%%%%%%%%%%%%%%%%%%%%%%%%%%%%%%%%%%%%%%%%%%%
%%%%%%%%%%%%%%%%%%%%%%%%%%%%%%%%%%%%%%%%%%%%%% Theory (Linear Model and WKB) %%%%%%%%%%%%%%%%%%%%%%%%%%%%%%%%%%%%%%%%%%%%%%

\section{Semi-analytical modelling of the 3D Batchelor vortex \label{section:Theory}}

In the following section, we linearize the governing equations of a perturbed Batchelor vortex in helical coordinates around the base flow of Eq.~(\ref{Batchelor_base}), while assuming that the flow is inviscid and $q$ can take any arbitrary value, to develop a 2D linear model enabling the investigation of the flow response to local disturbances. The PDEs of the linear model are then reduced to a less complicated system of ODEs by leveraging the WKB analysis, which is based on approximating perturbations as compact wave packets whose length scale is much smaller than that of the flow.  

\subsection{Linearized transport equations \label{section:linear_model}}

In order to develop a complexity-reduced, linear model for the 3D Batchelor vortex under consideration, we seek to rewrite its equations of motion in a 2D form via a transformation of the coordinate system from cylindrical coordinates $(r, \theta, z)$ to helical ones $(\rho, \chi, h)$. The unit vectors of the new coordinate system are specified as
\begin{equation}
	\bs{e}_{\rho} = \bs{e}_r \, , \quad \bs{e}_{\chi} = N^2\Big(\bs{e}_{\theta} - \frac{r}{L}\bs{e}_z \Big) \, , \quad \bs{e}_h = N^2 \Big( \bs{e}_z+ \frac{r}{L}\bs{e}_{\theta} \Big)   \, ,   \label{unit_vector} 
\end{equation}
with $2\pi L$ being the helix's pitch, and $N^2 = \big(1 + r^2/L^2\big)^{-1}$. The linear stability analysis on a 3D axisymmetric flow reveals that the most excited mode is concentrated around $r = r_0$, where $\mrm{d}/\mrm{d}r(\overline{u}_{\theta}/r - \overline{u}_z/L = 0)_{r = r_0}$ \citep{Leibovich1983}. Here, $L = -n/k_z$ is a constant, while $n$ and $k_z$ are the tangential and axial wavenumbers of the most amplified mode, respectively. $\overline{u}_{\theta}$ and $\overline{u}_{z}$ also represent the tangential and axial velocities of the base flow, respectively. The coordinates of the new system are provided by 
\begin{equation}
	\rho = r \, , \quad \chi = \theta - \frac{z}{L} \, , \quad h = N^2 z    \, .   \label{coordinates}
\end{equation}
As can be seen, the radial components of both coordinate systems are identical, and therefore will be used interchangeably. The derivatives with respect to $\chi$ can be calculated as functions of those with respect to $\theta$ and $z$ using the chain rule, such that
\begin{equation}
       \frac{\partial ( \ )}{\partial \theta} = \frac{\partial ( \ )}{\partial \chi}\frac{\partial \chi}{\partial \theta} + \frac{\partial ( \ )}{\partial z}\frac{\partial z}{\partial \theta}   \, .   \label{chain}
\end{equation}

The velocity field in helical coordinates is described by $\bs{u} = u_{\rho} \bs{e}_{\rho} + u_{\chi} \bs{e}_{\chi} + u_h \bs{e}_{h}$. The helical symmetry of such coordinate system yields
\begin{equation}
	\bs{e}_h \cdot \bnabla u_r = \bs{e}_h \cdot \bnabla u_{\chi} = \bs{e}_{h} \cdot \bnabla u_{h} = 0     \, ,   \label{symmetry}
\end{equation}
leading to
\begin{equation}
	u_{\rho} = u_r \, , \quad u_{\chi} = u_{\theta} - \frac{r}{L} u_z \, , \quad  u_h = u_z + \frac{r}{L} u_{\theta}    \, ,   \label{vel_field}
\end{equation}
and
\begin{equation}
	\omega_{\rho} = \omega_r \, , \quad \omega_{\chi} = \omega_{\theta} - \frac{r}{L} \omega_z \, , \quad  \omega_h = \omega_z + \frac{r}{L} \omega_{\theta}    \, .   \label{vort_field}
\end{equation}
The tangential and axial components of the velocity and vorticity fields can in turn be expressed in terms of the corresponding quantities in helical coordinates as
\begin{eqnarray}
	u_{\theta} = N^2\Big(u_{\chi} + \frac{r}{L} u_h\Big)  \, , \quad u_{z} = N^2\Big(u_{h} - \frac{r}{L} u_{\chi} \Big)   \, , \label{vel_field2} \\
	\omega_{\theta} = N^2\Big(\omega_{\chi} + \frac{r}{L} \omega_h\Big)  \, , \quad \omega_{z} = N^2\Big(\omega_{h} - \frac{r}{L} \omega_{\chi} \Big)  \label{vort_field2} \, . 
\end{eqnarray}
Eq. (\ref{symmetry}), representing the helical symmetry, is of certain significance, since it demonstrates that the velocity field can be fully described only by the two variables $r$ and $\chi$, instead of $r$, $\theta$ and $z$, which, subsequently, allows for recasting the flow dynamics in a 2D framework. The continuity equation in this framework reads
 \begin{equation}
 	\frac{\partial (ru_r)}{\partial r} + \frac{\partial u_{\chi}}{\partial \chi} = 0 \, ,  \label{continuity}
 \end{equation}
enabling the definition of the streamfunction $\psi$ as
 \begin{equation}
 	u_r = \frac{1}{r}\frac{\partial \psi}{\partial \chi} \, , \quad  u_{\chi} = -\frac{\partial \psi}{\partial r} \, .  \label{psi_def}
 \end{equation}
The velocity and vorticity fields can thus be formulated as
\begin{equation} 
 	\bs{u} = \bnabla \psi \times \bs{e}_h + u_h \bs{e}_h  \, , \quad \bs{\omega} = \bnabla u_h \times \bs{e}_h + \omega_h \bs{e}_h \, ,  \label{fields}
 \end{equation}
 while $\bs{\omega} = \bnabla \times \bs{u}$ has been employed, which in scalar form means
 \begin{equation} 
 	\omega_r = \frac{N^2}{r} \frac{\partial u_h}{\partial \chi} \, , \quad  \omega_{\theta} = \frac{N^2r}{L} \omega_h - N^2 \frac{\partial u_h}{\partial r} \, , \quad \omega_z = N^2 \omega_h + \frac{N^2r}{L} \frac{\partial u_h}{\partial r}   \, .  \label{omega_field}
 \end{equation}
 The relation $\bs{\omega} = \bs{\nabla} \times \bs{u}$ ultimately gives rise to the helical vorticity $\omega_{hel}$, calculated as  
 \begin{equation}
 	\omega_{hel} = N^2 \omega_h - \frac{2N^4}{L}u_h  \, , \label{w_hel_def}
 \end{equation}
 and connected to the streamfunction $\psi$ through
 \begin{equation}
 	\omega_{hel} = -\nabla^2 \psi  \, , \label{Poisson}
 \end{equation}
 where the Laplacian operator $\nabla^2$ is defined by
 \begin{equation}
	\nabla^2 \triangleq \frac{1}{r}\frac{\partial}{\partial r}\bigg(rN^2 \frac{\partial}{\partial r}\bigg) + \frac{1}{r^2}\frac{\partial^2 }{\partial \chi^2}  \, .  \label{Laplacian}
\end{equation}
 Note that $\nabla^2$ is the 2D Laplacian operator of the new coordinate system. The detailed derivation of the transport equations for the helical velocity $u_h$ and vorticity $\omega_{hel}$ can be found in \cite{Delbende2005}. In summary, the spatiotemporal evolution of $u_h$ and $\omega_{hel}$ is governed by
\begin{eqnarray}
 	\frac{\mrm{D} u_h}{\mrm{D} t}  &=&  0  \, , \label{uh_gov} \\
 	\frac{\mrm{D} \omega_{hel}}{\mrm{D} t} + \bigg[u_r \frac{\partial}{\partial r} + \Big(\frac{u_{\chi}}{r} + \frac{u_h}{L}\Big)\frac{\partial }{\partial \chi} \bigg]\frac{2N^4u_h}{L}  &=&  0   \, .   \label{wh_gov} 
\end{eqnarray} 
 where
\begin{equation}
 	\frac{\mrm{D}}{\mrm{D}t} = \frac{\partial}{\partial t} + u_r \frac{\partial}{\partial r} + \frac{u_{\chi}}{r}\frac{\partial}{\partial \chi}   \, .   \label{material_deriv} 
\end{equation}
    
We extend the decompositions of Eq.~(\ref{decompose}) to the variables in helical coordinates, and substitute them into Eqs. (\ref{uh_gov}) and (\ref{wh_gov}) to obtain
\begin{eqnarray}
 	\frac{\partial (\overline{u}_h + u_h')}{\partial t} + (\overline{u}_r + u_r') \frac{\partial (\overline{u}_h + u_h')}{\partial r} + \frac{\overline{u}_{\chi} + u_{\chi}'}{r}\frac{\partial(\overline{u}_h + u_h')}{\partial \chi}   &=&   0  \, , \label{uh_transport} \\
 \hspace{-0.33in}	\frac{\mrm{D} (\overline{\omega}_{hel} + \omega_{hel}')}{\mrm{D} t} + \bigg[(\overline{u}_r + u_r') \frac{\partial}{\partial r} + \bigg(\frac{\overline{u}_{\chi} + u_{\chi}'}{r} + \frac{\overline{u}_h + u_h'}{L}\bigg)\frac{\partial }{\partial \chi}\bigg]\frac{2N^4(\overline{u}_h + u_h')}{L}   &=&   0   \, .   \label{wh_transport} 
\end{eqnarray} 
The second-order terms can be neglected, and the transport equations of the base flow 
\begin{eqnarray}
 	\frac{\partial \overline{u}_h}{\partial t} + \overline{u}_r \frac{\partial \overline{u}_h}{\partial r} + \frac{\overline{u}_{\chi}}{r}\frac{\partial \overline{u}_h}{\partial \chi}   &=&   0  \, , \label{uh_transport_base} \\
 	\frac{\partial \overline{\omega}_{hel}}{\partial t} + \overline{u}_r \frac{\partial \overline{\omega}_{hel}}{\partial r} + \frac{\overline{u}_{\chi}}{r} \frac{\partial \overline{\omega}_{hel}}{\partial \chi} + \bigg[\overline{u}_r \frac{\partial}{\partial r} + \frac{\partial}{\partial r} \bigg(\frac{\overline{u}_{\chi}}{r} + \frac{\overline{u}_h}{L}\bigg)\frac{\partial }{\partial \chi}\bigg]\frac{2N^4 \overline{u}_h}{L}   &=&   0   \, ,   \label{wh_transport_base} 
\end{eqnarray} 
can be substracted from Eqs. (\ref{uh_transport}) and (\ref{wh_transport}) to arrive at
\begin{eqnarray}
 	\frac{\partial u_h'}{\partial t} + u_r' \frac{\partial \overline{u}_h }{\partial r} + \overline{u}_r \frac{\partial u_h'}{\partial r} + \frac{u_{\chi}'}{r}\frac{\partial\overline{u}_h}{\partial \chi} + \frac{\overline{u}_{\chi}}{r}\frac{\partial u_h'}{\partial \chi}   &=&   0  \, , \label{uh_transport_pert1} \\ 
 	\frac{\partial \omega_{hel}'}{\partial t} + u_r' \frac{\partial \overline{\omega}_{hel} }{\partial r} + \overline{u}_r \frac{\partial \omega_{hel}'}{\partial r} + \frac{u_{\chi}'}{r}\frac{\partial \overline{\omega}_{hel}}{\partial \chi} + \frac{\overline{u}_{\chi}}{r}\frac{\partial \omega_{hel}'}{\partial \chi}   &+&   \nonumber \\ \hspace{-0.36in}
 	\frac{2}{L}\bigg[\overline{u}_r \frac{\partial (N^4 u_h')}{\partial r} + u_r' \frac{\partial (N^4 \overline{u}_h)}{\partial r}\bigg] + \frac{2N^4}{L}\bigg[\bigg(\frac{\overline{u}_{\chi}}{r} + \frac{\overline{u}_h}{L}\bigg)\frac{\partial u_h'}{\partial \chi} + \bigg(\frac{u_{\chi}'}{r} + \frac{u_h'}{L}\bigg)\frac{\partial \overline{u}_h}{\partial \chi}\bigg]   &=&   0   \, .   \label{wh_transport_pert1} 
\end{eqnarray} 

In the ensuing analysis, we do not impose any constraints on the value of $q$. This, however, does not change the axisymmetry property and $\theta$-independence of the base flow, as $\overline{u}_z$ and other quantities of the base flow are solely functions of radius (cf. Eq.~(\ref{Batchelor_base})), which further suggests that the base flow is $\chi$-independent as well, since for an axisymmetric flow $\frac{\partial( \ )}{\partial \chi} = \frac{\partial ( \ )}{\partial \theta}$ (cf. Eq.~(\ref{chain})). We also recall that $\overline{u}_r = 0$, $\omega_{hel} = N^2 \omega_h - \frac{2N^4}{L} u_h = -\nabla^2 \psi$ and $r u_r = \frac{\partial \psi}{\partial \chi}$. These relations can be inserted into Eqs. (\ref{uh_transport_pert1}) and (\ref{wh_transport_pert1}) to further simplify them  to    
\begin{eqnarray}
 	\frac{\partial u_h'}{\partial t} + \frac{\eta}{L} \frac{\partial \psi}{\partial \chi} + \Omega_{\chi} \frac{\partial u_h'}{\partial \chi}   &=&   0  \, , \label{uh_transport_pert} \\
 	-\bigg(\frac{\partial}{\partial t} + \Omega_{\chi} \frac{\partial}{\partial \chi}\bigg) \nabla^2 \psi' + \frac{1}{r} \frac{\partial \psi'}{\partial \chi}\frac{\mrm{d} (N^2 \overline{\omega}_h)}{\mrm{d} r} + \frac{2\Omega N^2}{L} \frac{\partial u_h'}{\partial \chi}   &=&   0    \, ,   \label{wh_transport_pert} 
\end{eqnarray} 
where we have defined $\Omega_{\chi} \triangleq \frac{\overline{u}_{\chi}}{r}$, $\Omega \triangleq \frac{\overline{u}_{\theta}}{r}$ and $\eta \triangleq \frac{L}{r} \frac{\partial \overline{u}_h}{\partial r}$. 

\begin{figure}
	\centerline{\includegraphics[width=1.\textwidth]{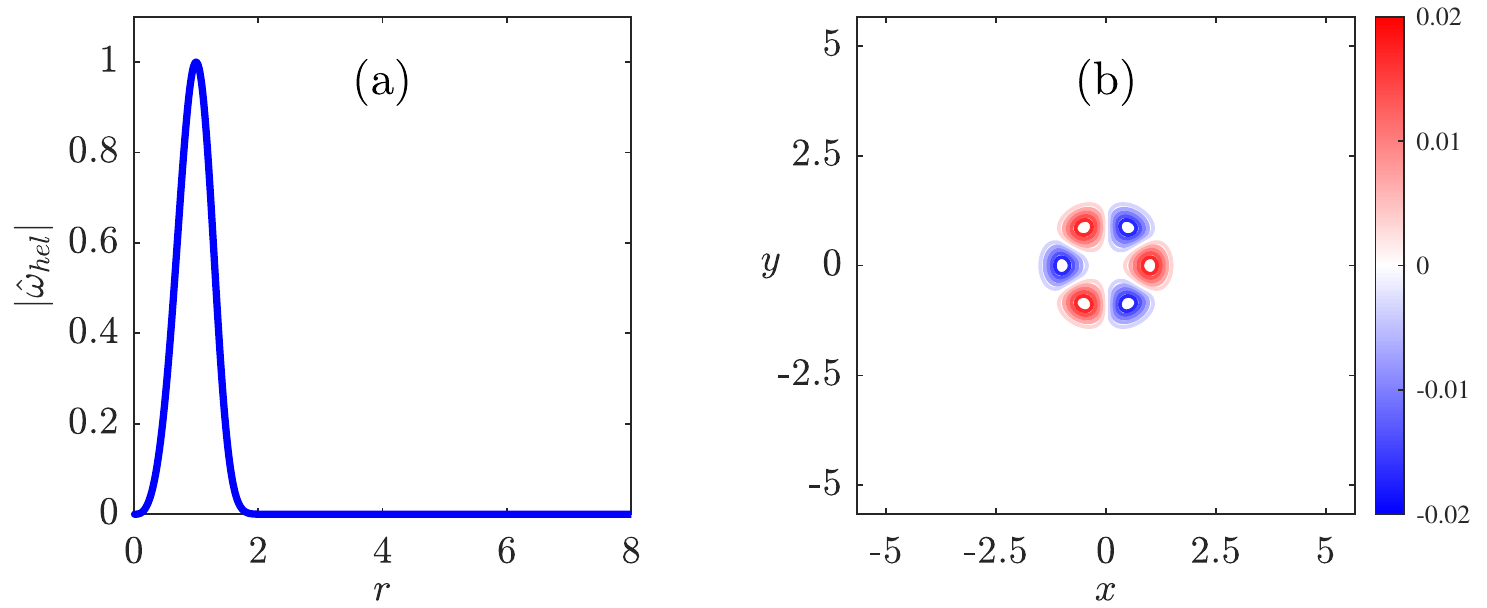}}
	\caption{\footnotesize The initial value of the perturbation helical vorticity (a) in the $\chi$-Fourier space, and (b) in the physical space (in the $z = 0$ plane), when $m_f = 2$. Note that since $u_h'$ is zero everywhere in the domain at $t = 0$, panel (b) also indicates the initial value of $\omega_z'$.} 
	\label{IC}
\end{figure}   

The Fourier transform of the linear model described by Eqs. (\ref{uh_transport_pert}) and (\ref{wh_transport_pert}) with respect to $\chi$ renders 
\begin{eqnarray}
 	\frac{\partial \hat{u}_{h,m}}{\partial t} + \imag m \frac{\eta}{L} \hat{\psi}_m + \imag m \Omega_{\chi} \hat{u}_{h,m}   &=&   0  \, , \label{uh_Fourier} \\
 	\bigg(\frac{\partial}{\partial t} + \imag m \Omega_{\chi} \bigg) \hat{\omega}_{hel,m} + \frac{\mrm{d} (N^2 \overline{\omega}_h)}{\mrm{d} r}\frac{\imag m \hat{\psi}_m}{r} + \imag m \frac{2\Omega N^2}{L} \hat{u}_{h,m}   &=&   0   \, ,    \label{wh_Fourier} 
\end{eqnarray}
where $\hat{ }$-sign signifies that a variable is in the $\chi$-Fourier space. In other words
\begin{equation}
	\hat{f}_m = \int_{-\infty}^{\infty} f(\chi) e^{-\imag m\chi} \mrm{d} \chi \, ,
\end{equation}
with $m$ being the $\chi$-wavenumber. Since $\chi = \theta - z/L$, $m$ can be related to the tangential and axial wavenumbers by
\begin{equation}
	m\chi = m\Big(\theta - \frac{z}{L}\Big) \Longrightarrow n = m \, , \ k_z = -\frac{m}{L} \, .  \label{wavenumbers}  
\end{equation}
Eqs.~(\ref{uh_Fourier}) and (\ref{wh_Fourier}) can be numerically solved along with
\begin{equation}
	\hat{\omega}_m = \bigg[\frac{m^2}{r^2}\hat{\psi} - \frac{1}{r}\frac{\partial}{\partial r}\bigg(N^2 r \frac{\partial \hat{\psi}}{\partial r} \bigg) \bigg]  \, , \label{closure}
\end{equation}
to determine how the initial perturbation prescribed by 
\begin{equation}
	\hat{\omega}_{h, m_f}(t = 0) = a_f \frac{r^{m_f}}{N^2}\exp\big[0.25m_f\big(1 - r^4\big)\big]   \, , \label{disturbance}
\end{equation}
evolves in time. Here, $a_f$ is the perturbation amplitude in the $\chi$-Fourier space, and $m_f$ shows the forced $\chi$-wavenumber. The computational domain is assumed to be sufficiently large so that its boundaries remain undisturbed. The perturbation of Eq. (\ref{disturbance}) can be expressed in the physical space as 
\begin{equation}
	\omega_h'(t = 0) = a_p \frac{r^{m_f}}{N^2}\exp\big[0.25m_f\big(1 - r^4\big)\big] \cos(m_f \chi)   \, , \label{disturbance_phys}
\end{equation}
where $a_p$ is the perturbation amplitude in the physical space. In this paper, we take $a_f = 1$, and we choose 50 Fourier modes or collocation points (i.e., $N_m = 50$) for the numerical solution of Eqs.~(\ref{uh_Fourier}) and (\ref{wh_Fourier}), yielding $a_p = 0.02$, as $a_p = a_f/N_m$. Since $\omega_{r}$ and $\omega_{\chi}$ are left unperturbed at $t = 0$, the vorticity field in cylindrical coordinates can be calculated using Eq. (\ref{vort_field2})  
\begin{eqnarray}
	\omega_r'(t = 0)  &=&  0  \, , \nonumber \\
	\omega_{\theta}'(t = 0)  &=&  a_p \frac{r^{m_f+1}}{L} \exp\big[0.25m_f\big(1 - r^4\big)\big] \cos(m_f \chi) \, , \label{disturbance_cyl} \\
	\omega_z'(t = 0)  &=&  a_p r^{m_f} \exp\big[0.25m_f\big(1 - r^4\big)\big] \cos(m_f \chi) \, ,  \nonumber 
\end{eqnarray}
which in the Cartesian coordinates of the DNS solver described in \S \ref{section:Setup} read
\begin{eqnarray}
	\omega_x'(t = 0)  &=&  -a_p \frac{r^{m_f+1}}{L} \exp\big[0.25m_f\big(1 - r^4\big)\big] \cos(m_f \chi)\sin(\theta)  \, , \nonumber \\
	\omega_y'(t = 0)  &=&  a_p \frac{r^{m_f+1}}{L} \exp\big[0.25m_f\big(1 - r^4\big)\big] \cos(m_f \chi)\cos(\theta) \, , \label{disturbance_cart} \\
	\omega_z'(t = 0)  &=&  a_p r^{m_f} \exp\big[0.25m_f\big(1 - r^4\big)\big] \cos(m_f \chi) \, .  \nonumber
\end{eqnarray}
Furthermore, because $\omega_{\theta}' = (N^2r/L) \omega_h'$ and $\omega_z' = N^2 \omega_h'$ at $t = 0$, it can be deduced from Eq. (\ref{omega_field}) that initially $\frac{\partial u_h'}{\partial r} = 0$, leading to $u_h' = 0$. The latter holds, since $u_h'$ has to vanish as $r \rightarrow \infty$. This result along with Eq.~(\ref{disturbance}) can be replaced into Eq.~(\ref{w_hel_def}) to arrive at
\begin{equation}
 	\hat{\omega}_{hel, m_f}(t = 0) = N^2 \hat{\omega}_{h, m_f}(t = 0) = r^{m_f}\exp\big[0.25m_f\big(1 - r^4\big)\big]   \, . \label{disturbance_hel}
\end{equation}
The initial perturbation helical vorticity $\omega_{hel}' (t = 0)$ is shown in Fig.~\ref{IC}, in both the $\chi$-Fourier and physical spaces.

\subsection{Wenzel-Kramers-Brillouin analysis \label{section:WKB}}

The WKB analysis hinges on modelling the perturbation variables as compact wavepackets, i.e., the perturbation streamfunction $\psi'$ and the perturbation helical velocity $u_h'$ can be approximated as
\begin{eqnarray}
    	\psi'(r, \chi, t)   & \approx &   K_1(r, t) \lambda(\chi) \, , \label{psi_pert}  \\
    	u_h'(r, \chi, t)    & \approx &   K_2(r, t) \lambda(\chi) \, , \label{uh_pert}  
\end{eqnarray}
where
\begin{equation}
    	\lambda(\chi) = \exp(\imag m \chi)  \Longrightarrow \frac{\partial \psi'}{\partial \chi} = \imag mK_1 \lambda \, , \frac{\partial u_h'}{\partial \chi} = \imag m K_2 \lambda \, ,  \label{theta_part}
\end{equation}
and
\begin{eqnarray}
    	K_1(r, t)   &=&   A_1(t) \exp\big\{\imag [k(t)(r_1 - R) - \Lambda_1(t)] \big\} \, ,   \label{rt_part1} \\ 
    	K_2(r, t)   &=&   A_2(t) \exp\big\{\imag [k(t)(r_2 - R) - \Lambda_2(t)] \big\} \, .   \label{rt_part2}  
\end{eqnarray}
Here, $r_1$ and $r_2$ are the radial locations of the perturbation waves of the streamfunction and helical velocity, respectively, and $R$ is the radius at which the initial perturbation is applied, which, subsequently, is also the radius around which the solution is sought. $k$, $A$ and $\Lambda$ also represent the radial wavenumber, the wave amplitude and the wave phase, respectively. The time derivatives of $K_1$ and $K_2$ can be evaluated as   
\begin{eqnarray}
    	\dot{K}_1   &=&   K_1 \frac{\dot{A_1}}{A_1} + K_1\big[\imag \dot{k}(r_1 - R) - \imag \dot{\Lambda}_1\big] \, ,  \nonumber  \\
    	\dot{K}_2   &=&   K_2 \frac{\dot{A_2}}{A_2} + K_2\big[\imag \dot{k}(r_2 - R) - \imag \dot{\Lambda}_2\big] \, ,  \label{time_derivative}
\end{eqnarray}
where overdot shows the temporal derivative. 

\vspace{0.075in}

The WKB analysis relies on two key assumptions:
\begin{enumerate}   
	\vspace{0.025in}
	\item The wave is tightly wound (compact), i.e., $kR \gg 1$.	
	\item The length scale of the wave propagation is much smaller than that of the flow field, therefore
	\begin{eqnarray}
		r - R   &=&   \delta r \ll 1 \, , \nonumber \\  
		\frac{1}{r}   &=&   \frac{1}{R}\Big(1 - \frac{\delta r}{R} + \cdots\Big) \, , \label{linear_expansion}
	\end{eqnarray}
	where $r$ can be replaced by the radial location of either wave. Consequently,
	\begin{equation}
		\frac{1}{r} \frac{\partial}{\partial r}\bigg[N^2 r\frac{\partial ( \ )}{\partial r}\bigg] \approx \frac{1}{R} \frac{\partial}{\partial r}\bigg[R N^2(r = R)\frac{\partial ( \ )}{\partial r}\bigg] = N_0^2 \frac{\partial^2 ( \ )}{\partial r^2} \, ,    \label{radial_derivative_approx}
	\end{equation}
	where $N_0 = N(r = R)$. Note also that
	\begin{equation}
		\frac{1}{r} \frac{\mrm{d} (rN^2)}{\mrm{d} r}\bigg|_{r = R} \frac{\partial ( \ )}{\partial r} \sim \mathcal{O}(\delta r) \ll N^2 \frac{\partial^2 ( \ )}{\partial r^2} \sim \mathcal{O}(\delta r^2) \, .    \label{derivative_approx_note}
	\end{equation}
\end{enumerate}
Eqs. (\ref{radial_derivative_approx}) and (\ref{derivative_approx_note}) thus result in  
\begin{equation}
	\frac{1}{r} \frac{\partial}{\partial r}\bigg(rN^2 \frac{\partial K_1}{\partial r}\bigg) \approx N_0^2 \frac{\partial^2 K_1}{\partial r^2} = -k^2 N_0^2 K_1 \, .    \label{radial_derivative}
\end{equation}
We also define
\begin{equation}
	 C \triangleq \frac{\mrm{d} (N^2 \overline{\omega}_h)}{\mrm{d} r}  \, . \label{Def}  
\end{equation}

The linear expansion of $\Omega_{\chi}$ and $\Omega$ around $R$ yields
\begin{eqnarray}
	\Omega_{\chi}   &\approx&   \Omega_{\chi, 0} + \frac{\mrm{d} \Omega_{\chi}}{\mrm{d} r} \bigg|_{r = R} \delta r \, ,  \nonumber \\ 
	\Omega   &\approx&   \Omega_{\chi,0} + \frac{\mrm{d} \Omega}{\mrm{d} r} \bigg|_{r = R} \delta r \, ,  \label{linear_exp}
\end{eqnarray}
where the $0$-subscript indicates that a variable has been evaluated at $r = R$. We then substitute Eqs. (\ref{time_derivative}), (\ref{radial_derivative}) and (\ref{linear_exp}) into Eq. (\ref{uh_transport_pert}) to arrive at
\begin{eqnarray}
	\frac{\dot{A}_2}{A_2} K_2 \lambda + \imag (\dot{k}\delta r - \dot{\Lambda}_2)K_2\lambda + \frac{\imag m \eta_0}{L} K_1 \lambda + \imag m(\Omega_{\chi,0} + \Omega_{\chi,0}' \delta r)K_2 \lambda = 0 \, ,   \label{WKB_uh1} 
\end{eqnarray}
and into Eq. (\ref{wh_transport_pert}) to derive
\begin{eqnarray}
	2k\dot{k}N_0^2K_1 \lambda + \bigg(k^2N_0^2 + \frac{m^2}{R^2}\bigg)\bigg[\frac{\dot{A}_1}{A_1} + \imag \big(\dot{k}\delta r -  \dot{\Lambda}_1\big)\bigg]K_1 \lambda + \nonumber \\ \imag m(\Omega_{\chi,0} + \Omega_{\chi,0}' \delta r) \bigg(k^2N_0^2 + \frac{m^2}{R^2} \bigg)K_1 \lambda + \frac{\imag m C_0}{R} K_1 \lambda + \imag m \frac{2\Omega_0N_0^2}{L^2} K_2 \lambda = 0   \, . \label{WKB_psi1}
\end{eqnarray}
Eqs. (\ref{WKB_uh1}) and (\ref{WKB_psi1}) can respectively be divided by $K_2 \lambda$ and $K_1 \lambda$ to obtain
\begin{eqnarray}
	\frac{\dot{A}_2}{A_2} + \imag(\dot{k}\delta r - \dot{\Lambda}_2) + \frac{\imag m \eta_0}{L}\frac{A_1}{A_2} (\cos \phi - \imag \sin \phi) + \imag m(\Omega_{\chi,0} + \Omega_{\chi,0}' \delta r) = 0  \, ,  \label{WKB_uh2} 
\end{eqnarray}
and
\begin{eqnarray}
\hspace{-0.22in}	2N_0^2k\dot{k} + \bigg(k^2N_0^2 + \frac{m^2}{R^2}\bigg)\bigg[\frac{\dot{A}_1}{A_1} + \imag(\dot{k}\delta r - \dot{\Lambda}_1)\bigg] &+&  \imag m(\Omega_{\chi,0} + \Omega_{\chi,0}' \delta r) \bigg(k^2N_0^2 + \frac{m^2}{R^2} \bigg) + \nonumber \\  \frac{\imag mC_0}{R} &+& \imag m \frac{A_2}{A_1}\frac{2\Omega_0N_0^2}{L^2} (\cos \phi + \imag \sin \phi) = 0   \, , \label{WKB_psi2}
\end{eqnarray}
where $\phi = \Lambda_1 - \Lambda_2$. The trigonometric functions are produced by $K_2/K_1 = \exp[\imag(\Lambda_1 - \Lambda_2)]$ (or $K_1/K_2$). We then separate the real and imaginary parts of the zeroth- and first-order terms of Eqs. (\ref{WKB_uh2}) and (\ref{WKB_psi2}) to acquire
\begin{eqnarray}
	\frac{\dot{A_2}}{A_2} + \frac{mA_1}{LA_2} \eta_0 \sin \phi  &=&  0 \, , \label{uh_real} \\
	\dot{\Lambda}_2 - m\Omega_{\chi,0} - \frac{mA_1}{LA_2}\eta_0 \cos \phi  &=&  0 \, ,\label{uh_imag}\\
	\dot{k} + m\Omega_{\chi,0}'  &=&  0 \label{uh_dr_imag}\, ,  
\end{eqnarray}
and
\begin{eqnarray}
	2N_0^2k\dot{k} + \bigg(k^2 N_0^2 + \frac{m^2}{R^2}\bigg)\frac{\dot{A}_1}{A_1} - \frac{2m\Omega_0 N_0^2}{L}\frac{A_2}{A_1} \sin \phi  &=&  0 \, , \label{wh_real} \\
	(m\Omega_{\chi,0} - \dot{\Lambda}_1)\bigg(k^2 N_0^2 + \frac{m^2}{R^2}\bigg) + \frac{m}{R}C_0 + \frac{2m\Omega_0 N_0^2}{L}\frac{A_2}{A_1} \cos \phi  &=&  0 \, , \label{wh_imag}\\
	(\dot{k} + m\Omega_{\chi,0}')\bigg(k^2N_0^2 + \frac{m^2}{R^2}\bigg)   &=&   0  \label{wh_dr_imag} \, .  
\end{eqnarray}

Given that $k^2N_0^2 + m^2/R^2 > 0$ for all $t \geq 0$, Eq.~(\ref{wh_dr_imag}) is satisfied if and only if $\dot{k} + m\Omega'_{\chi, 0} = 0$, which renders Eqs.~(\ref{uh_dr_imag}) and (\ref{wh_dr_imag}) identical, thereby making the number of unknowns and equations in the ODE system of Eqs.~(\ref{uh_real})--(\ref{wh_dr_imag}) equal. The radial wavenumber thus varies with time as $k(t) = k_0 - m\Omega'_{\chi,0}t$. Since the radial length scale of asymmetries used in this study (Fig.~\ref{IC}) is roughly about the initial core radius $r_{co}$ in dimensional form and, consequently, about 1, when variables are nondimensionalized, $k_0 = 1$ is a natural choice \citep{Montgomery1997}. Furthermore, the crucial assumption of the WKB analysis that $kR \gg 1$ requires limiting the scope of analysis to $R \geq 1$. Hence, in the remainder of the paper, we take $R = 1$.

A forward Euler scheme with the time step $\Delta t = 0.001$ is adopted to numerically integrate Eqs.~(\ref{uh_real}), (\ref{uh_imag}), (\ref{wh_real}) and (\ref{wh_imag}), and to evaluate the temporal variation of the remaining four unknowns $A_1$, $A_2$, $\Lambda_1$ and $\Lambda_2$. This then allows for the calculation of the radial group velocities $C_{g1}$ and $C_{g2}$ as
\begin{equation}
	C_{g1} = \frac{\partial \dot{\Lambda}_1}{\partial k}  \, , \quad C_{g2} = \frac{\partial \dot{\Lambda}_2}{\partial k}  \, , \label{group_vel} 
\end{equation} 
which in turn provide the radial locations of the helical vorticity and velocity perturbation waves via  
\begin{equation}
	r_1(t) = R + \int_0^t C_{g1}(\tau) \mrm{d}\tau \, , \quad r_2(t) = R + \int_0^t C_{g2}(\tau) \mrm{d}\tau \, .  \label{wave_location} 
\end{equation} 
The amplitude $A_{\omega}$ of the helical vorticity perturbation wave can also be estimated as $(m^2/R^2 + k^2N_0^2)A_1$.

\citet{Montgomery1997} primarily attributed the restricted radial propagation of disturbances in 2D trailing vortices to the rapid growth of the radial wavenumber $k(t)$, corresponding to constantly increasing shearing effects. These shearing effects will eventually dominate the flow dynamics, halting the outward propagation of the perturbation waves. \citet{Pradeep2006} have also identified a shearing mechanism caused by the differential advection of axial vorticity via mean swirl, which ultimately alters the `tilt' of streamlines so that the production of $u_r' u'_{\theta}$ is steadily reduced until it becomes negative, resulting in the arrest of transient growth in 2D vortex-dominated flows. These findings, however, are not consistent with the physics observed for 3D flows, as will be discussed in the next section. 

%%%%%%%%%%%%%%%%%%%%%%%%%%%%%%%%%%%%%%%%%%%% Results of the Linear Model and WKB Analysis %%%%%%%%%%%%%%%%%%%%%%%%%%%%%%%%%%%%%%%%%%%%
%%%%%%%%%%%%%%%%%%%%%%%%%%%%%%%%%%%%%%%%%%%% Results of the Linear Model and WKB Analysis %%%%%%%%%%%%%%%%%%%%%%%%%%%%%%%%%%%%%%%%%%%%

\section{Results of the linear analysis \label{section:Results}}

In this section, the results of the linear model developed in the previous section are compared with those of the WKB analysis for two scenarios of $q \rightarrow \infty$ and finite swirl numbers, separately. The section is concluded by finding an analytical relation from the WKB ODEs to determine conditions under which this analysis also envisages the unbounded propagation of asymmetries.

\subsection{Limiting case $q \rightarrow \infty$ \label{section:q_infinity}}

\begin{figure}
	\centerline{\includegraphics[width=1.\textwidth]{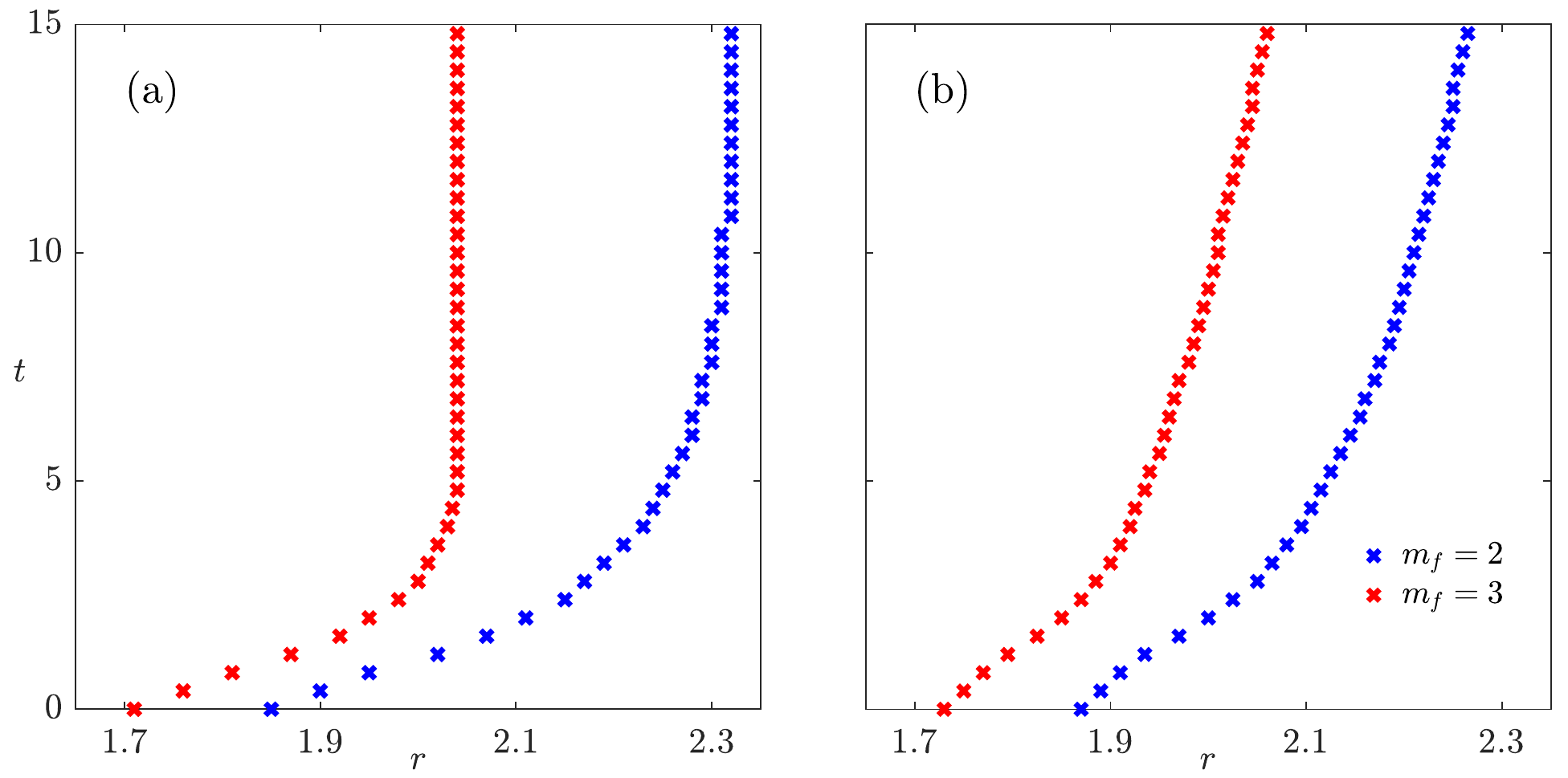}}
	\caption{\footnotesize The radial propagation of disturbances as functions of time for (a) a 2D Batchelor vortex, and (b) a 3D Batchelor vortex with $L = 2$ and $q \rightarrow \infty$. Note that in 2D flows $\omega_z$ is perturbed.} 
	\label{rt_plot}
\end{figure}

\begin{figure}
	\centerline{\includegraphics[width=1.\textwidth]{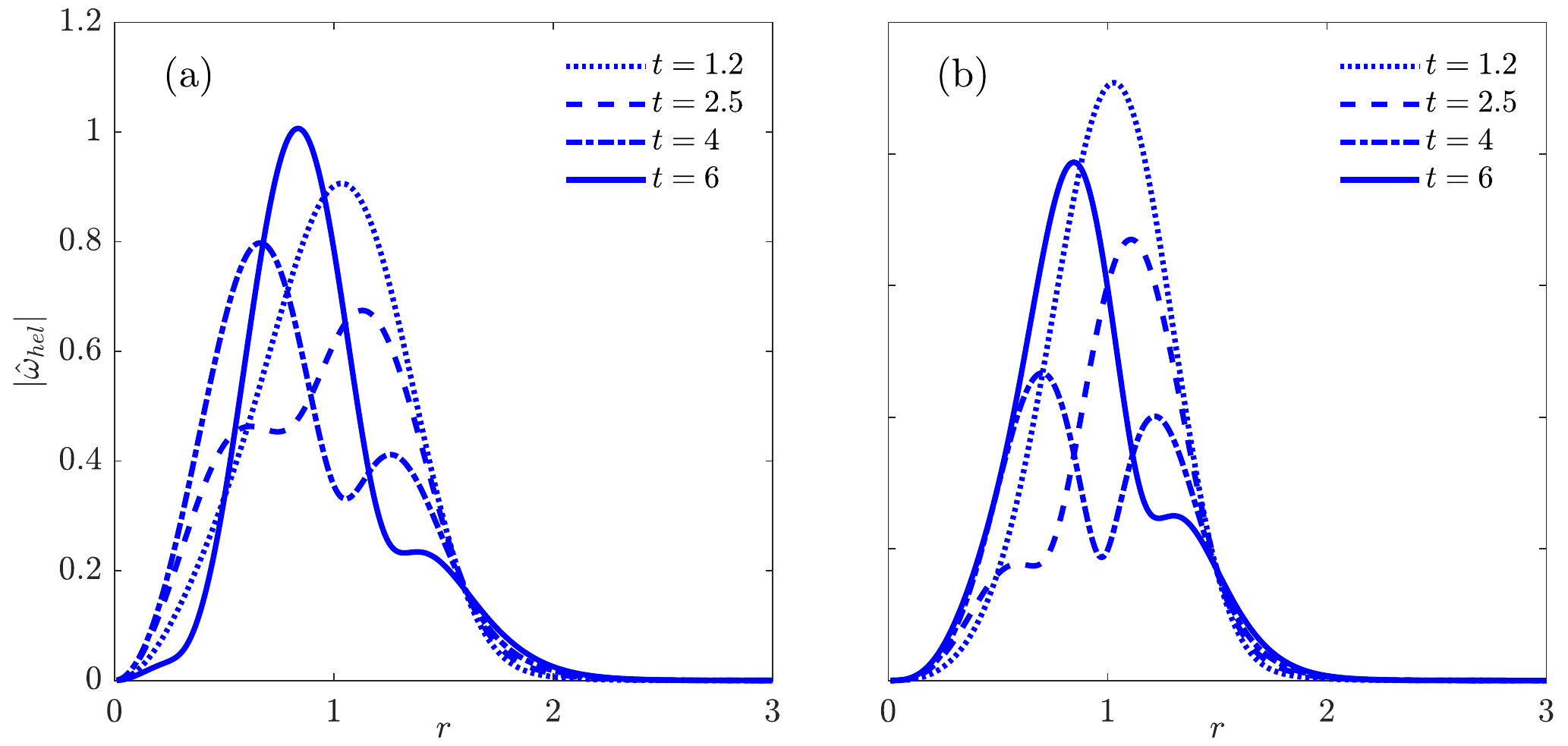}}
	\caption{\footnotesize The spatiotemporal evolution of the helical vorticity perturbation waves as functions of time when (a) $m_f = 2$, and (b) $m_f = 3$. In both panels, $L = 2$ and $q \rightarrow \infty$.} 
	\label{wave_evolution}
\end{figure}

\begin{figure}
	\centerline{\includegraphics[width=1.\textwidth]{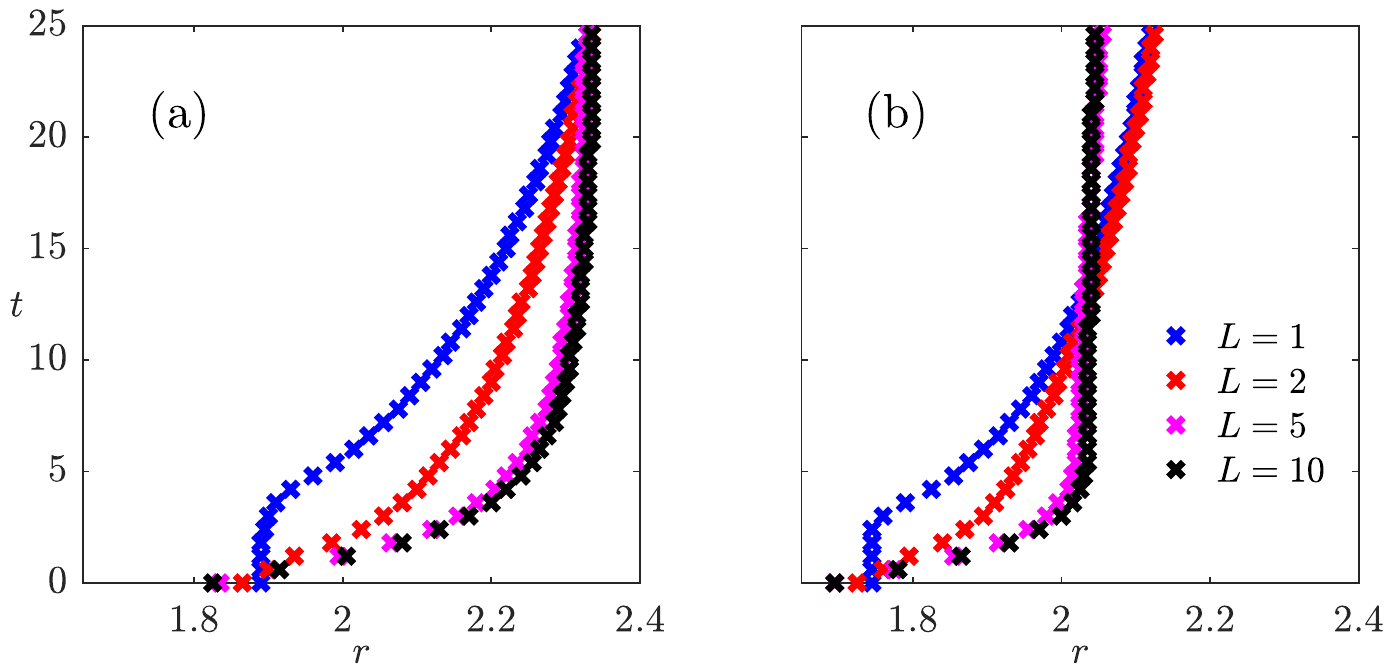}}
	\caption{\footnotesize The radial progression of perturbations with time at different values of $L$ for (a) $m_f = 2$, and (b) $m_f = 3$, provided by the numerical solution of Eqs.~(\ref{uh_Fourier}) and (\ref{wh_Fourier}). For all cases, $q \rightarrow \infty$.} 
	\label{L_effect_ode}
\end{figure}

\begin{figure}
	\centerline{\includegraphics[width=1.\textwidth]{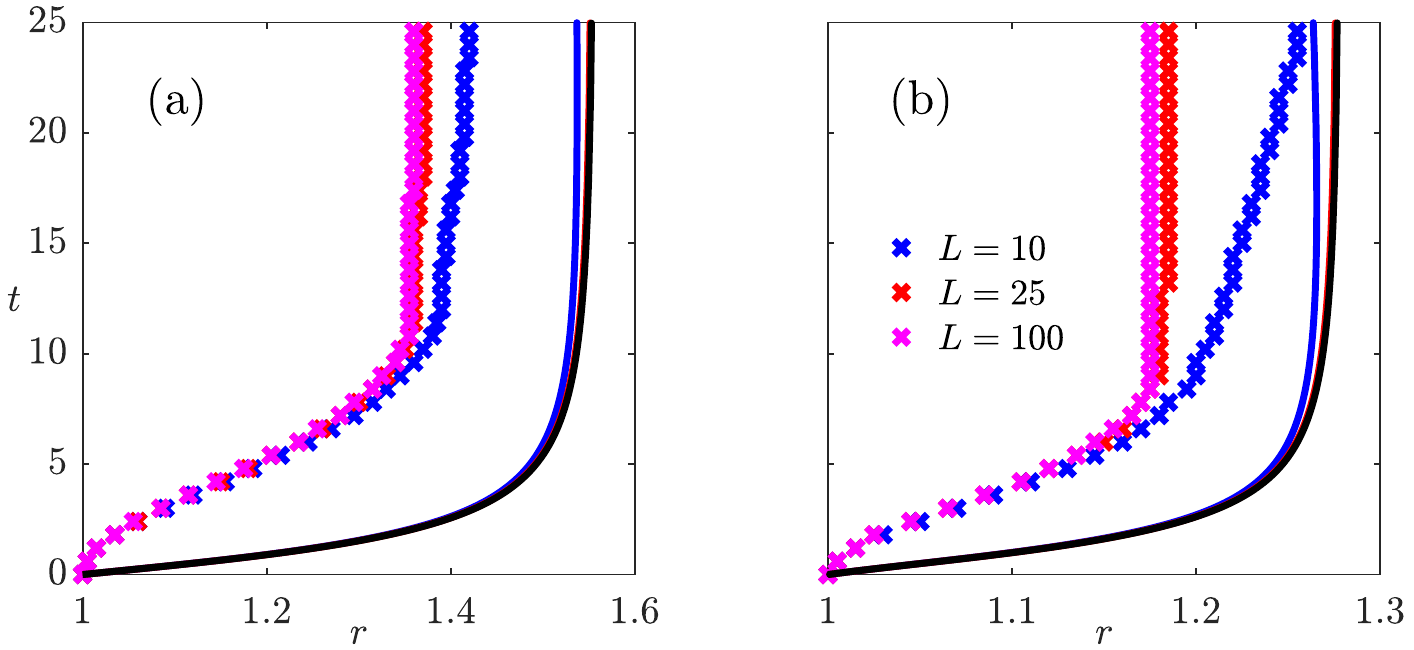}}
	\caption{\footnotesize Comparison between the predictions of the linear model of Eqs.~(\ref{uh_Fourier}) and (\ref{wh_Fourier}) (discrete crosses) and the results of the WKB analysis (solid lines) at large values of $L$, when (a) $m_f = 2$, and (b) $m_f = 3$. In both panels, $\overline{u}_z = 0$. Here, for the sake of a fair comparison between the linear model and the WKB analysis, the wave's peak (instead of its front) has been tracked in time, which is attainable, since for large values of $L$, the wave's peak either does not vanish or disappears after a very long time. Note also that the WKB results for $L = 25$ and $L = 100$ have collapsed on each other, since when $L$ is very large, the WKB predictions are nearly insensitive to $L$.} 
	\label{L_effect_compare}
\end{figure}

Before proceeding to the results for arbitrary swirl numbers, it is instructive to first focus on the limit $q \rightarrow \infty$, or, alternatively, $\overline{u}_z = 0$, where the Batchelor vortex is simplified to the Lamb-Oseen vortex \citep{Green1995}, due to its relative simplicity and its commonalities with cases in which $q$ is finite. In this scenario, $\overline{u}_{\chi} = \overline{u}_{\theta}$ and $\overline{u}_h = \frac{r}{L} \overline{u}_{\theta}$, consequently rendering $\Omega_{\chi} = \Omega = \overline{u}_{\theta}/r$ and $\eta = \overline{\omega}_z$. The latter relation is obtained using $\overline{\omega}_z = (1/r)[\partial(r \overline{u}_{\theta})/\partial r]$. The linear model of Eqs.~(\ref{uh_Fourier}) and (\ref{wh_Fourier}) can then be numerically integrated in time, where the streamfunction and the helical vorticity are linked via the closure equation (\ref{closure}), to determine the spatiotemporal evolution of the perturbations added to $\omega_{hel}$, as detailed in the previous section. The temporal variation of the front location of the perturbation waves, displayed in Fig. \ref{rt_plot}, is then measured by tracking the largest $r$ for which $\lvert \hat{\omega}_{hel,_m} \rvert > 0.01$. This method has been chosen in lieu of tracking the wave's peak, since for small and moderate values of $L$, the original peak gradually vanishes, while being replaced by a trailing extremum (Fig.~\ref{wave_evolution}). Note that in 2D and 3D flows, the perturbations are added to $\omega_z$ and $\omega_{hel}$, respectively. As can be observed in Fig.~\ref{rt_plot}, the outward propagation of disturbances in 2D cases is restrained, whereas in 3D flows, disturbances can travel in radial direction unboundedly. In fact, the linear model of (\ref{uh_Fourier}) and (\ref{wh_Fourier}) exhibits an unrestrained radial propagation for perturbation waves, so long as $L$ is not too large, i.e., when $L \lesssim 10$ (Figs.~\ref{L_effect_ode} and \ref{L_effect_compare}). The WKB analysis, on the other hand,  predicts that, for all values of $L$, the outward propagation of perturbations waves is contained. This asymptotic behaviour is anticipated, since for $q \rightarrow \infty$, $\Omega'_0 = \Omega'_{\chi,0}$ is finite and fairly large. One should, however, notice that, for small or modest values of $L$, it may take a relatively long time for the asymptotic behaviour of the WKB predictions to emerge, as the WKB analysis demonstrates a transient inward movement for the waves, interestingly occurring around the same period during which the wave's peak is forecast to disappear by the linear model. When $L \lesssim 1.1$, this inward propagation results in $r_1 < 0$, yielding the WKB results invalid.      

\begin{figure}
	\centerline{\includegraphics[width=1.\textwidth]{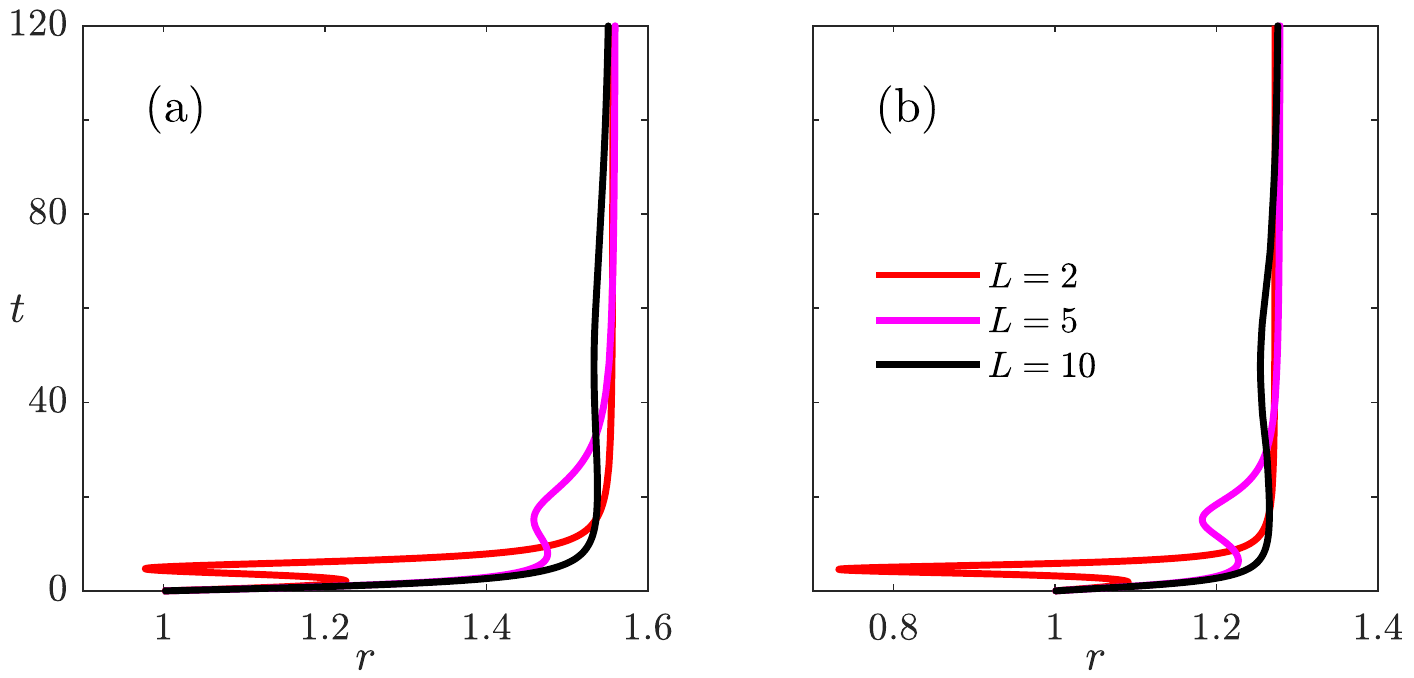}}
	\caption{\footnotesize The radial propagation of disturbances with time at different values of $L$ for (a) $m_f = 2$, and (b) $m_f = 3$, rendered by the WKB analysis. For all cases, $q \rightarrow \infty$. Note that the results for $L = 1$ is not presented, since the WKB analysis predicts $r_1 < 0$ when $L \lesssim 1.1$. The long time axis in both panels ensures that disturbances reach their asymptotic values.} 
	\label{L_effect_wkb}
\end{figure}

It is also informative to examine the impact of $L = -n/k_z$ on the results of the linear model and the WKB analysis. In the limit $L \rightarrow \infty$, Eqs.~(\ref{uh_transport_pert}) and (\ref{wh_transport_pert}) will be decoupled, and given that $\overline{u}_{\theta}$ is the only nonzero component in the velocity field of the base flow (since $q \rightarrow \infty$), Eq.~(\ref{wh_transport_pert}) will recover the linearized vorticity transport equation of the 2D Batchelor vortex. This suggests that as $L$ grows, the outward progression of the perturbation waves should steadily become more confined, and when $L$ is sufficiently large, the asymptotic behaviour observed for 2D flows should be restored. Figures \ref{L_effect_ode} and \ref{L_effect_compare} corroborate this insight by revealing that when $L \gtrsim 10$, one can reasonably expect a restrained radial propagation of disturbances. The agreement between the predictions of the linear model and the WKB analysis also improves with the growth of $L$ (Fig.~\ref{L_effect_compare}), as the fundamental assumption of the WKB analysis requiring a tightly-wound wave becomes progressively more well-founded. On the contrary, as $L$ diminishes, the perturbation waves can travel unboundedly outward, indicated by the finite value of $\mrm{d}t/\mrm{d}r$ at large times. A smaller value of $L$ corresponds to a smaller slope in the radius-time plot or, equivalently, a larger radial group velocity $C_{g1}$, at large values of $t$, which in turn reassures the farther and faster propagation of disturbances. This is in accordance with the findings of \citet{Pradeep2006} that, at a fixed azimuthal wavenumber, the growth of a perturbation mode is accelerated with the increase of $k_z$ or, analogously, with the decrease of $L$.     

\subsection{Finite swirl numbers \label{section:q_finite}}

\begin{figure}
	\centerline{\includegraphics[width=1.\textwidth]{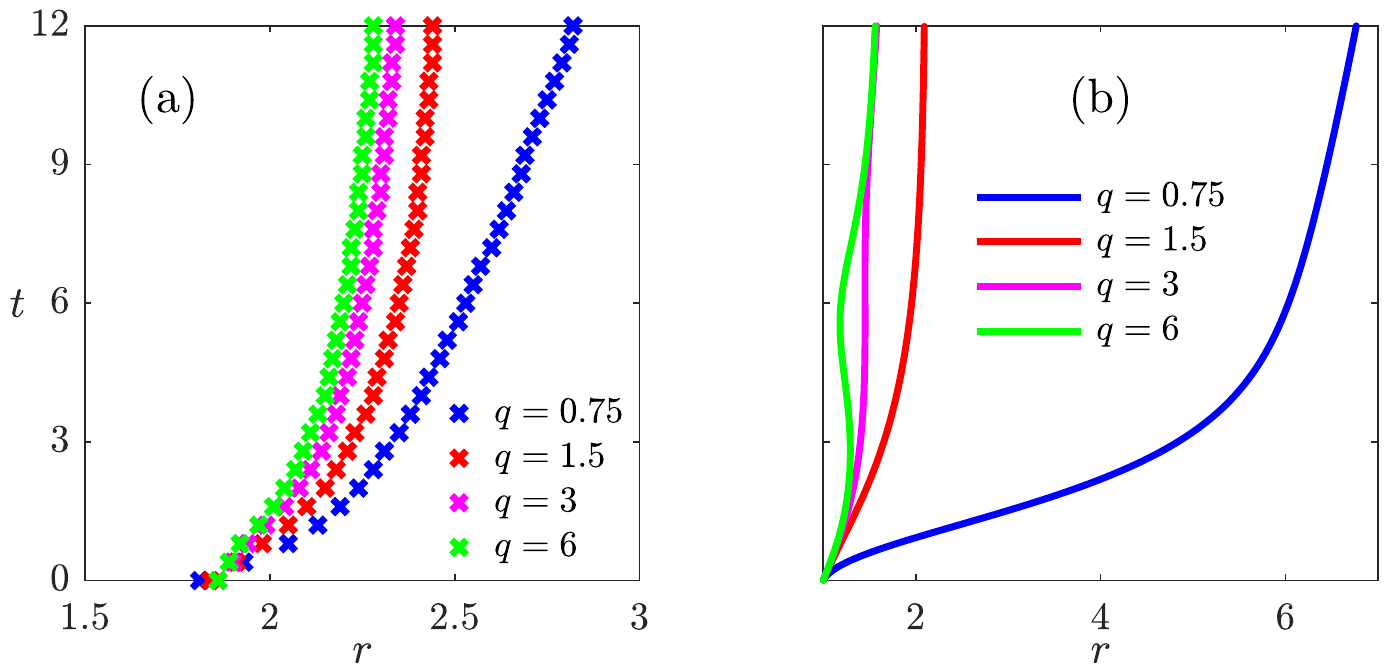}}
	\caption{\footnotesize The results of (a) the linear model of Eqs.~(\ref{uh_Fourier}) and (\ref{wh_Fourier}), and (b) the WKB analysis, for the radial propagation of disturbances at different values of $q$, when $m_f = L = 2$.} 
	\label{q_effect_rt}
\end{figure}

\begin{figure}
	\centerline{\includegraphics[width=.85\textwidth]{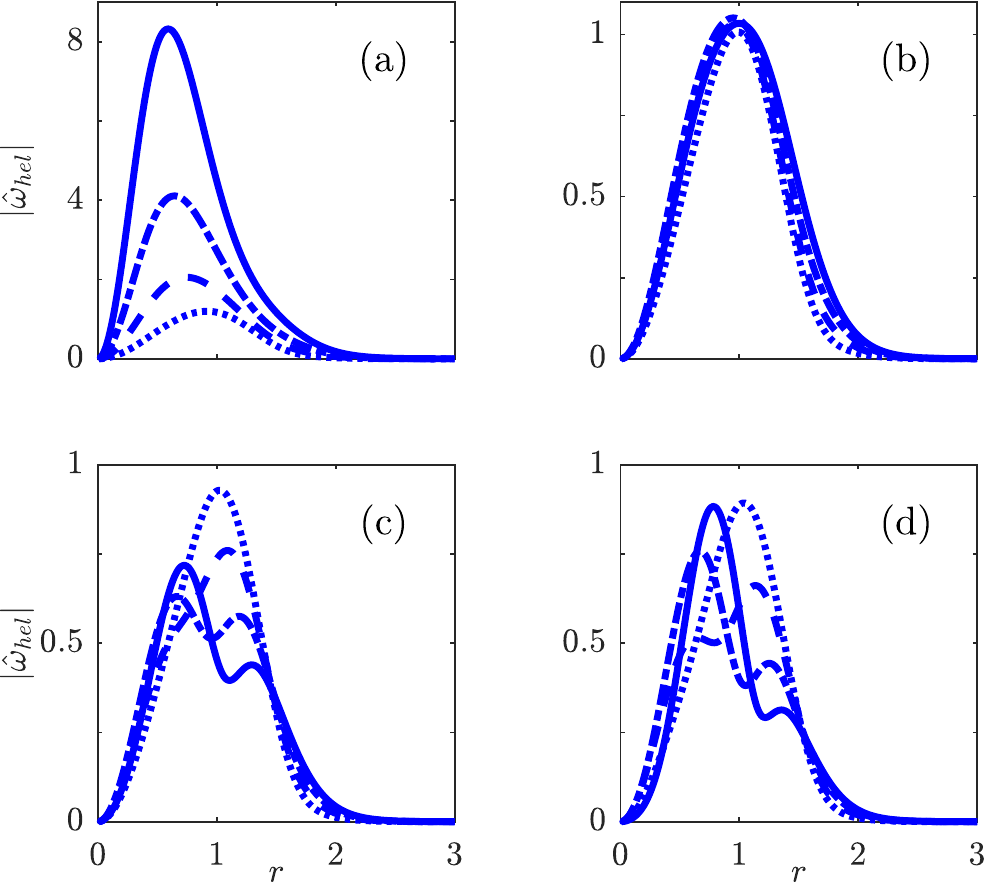}}
	\caption{\footnotesize The spatiotemporal evolution of the helical vorticity perturbation waves for (a) $q = 0.75$, (b) $q = 1.5$, (c) $q = 3$, and (d) $q = 6$, predicted by the numerical solution of Eqs.~(\ref{uh_Fourier}) and (\ref{wh_Fourier}), and shown at $t = 1.5$ (dotted lines), $t = 3$ (dashed lines), $t = 4.5$ (dash-dotted lines), and $t = 6$ (solid lines). In all examples, $m_f = L = 2$.} 
	\label{wave_shape}
\end{figure}

In this section, we return our attention to cases with arbitrary swirl numbers. Following the same procedure as the preceding section, the linear model of Eqs.~(\ref{uh_Fourier}) and (\ref{wh_Fourier}) can be numerically solved, and the front locations of the perturbation waves, displayed in Fig. \ref{q_effect_rt}(a), can be quantified as functions of time. As shown in this figure, the outward propagation of perturbation waves is suppressed as $q$ increases. Analogously, a smaller swirl number causes a less confined and more rapid progression of disturbances, indicated by lower slopes in the radius-time plots, which correspond to the higher values of $C_{g1} = \mrm{d}r/\mrm{d}t$. This trend is not monotonic, however, and reverses at $q \approx 0.3$. Furthermore, we find that for sufficiently small values of $q$, the amplitude of a disturbance continually and rapidly grows with time, instantly after its introduction to the flow (Fig.~\ref{wave_shape}). These observations are in accordance with the findings of existing studies, reporting that helical instabilities, capable of intensifying local disturbances and facilitating their outward propagation, emerge for $q \lesssim 1.5$ \citep{Lessen1974, Duraisamy2008}.  

\begin{figure}
	\centerline{\includegraphics[width=1.\textwidth]{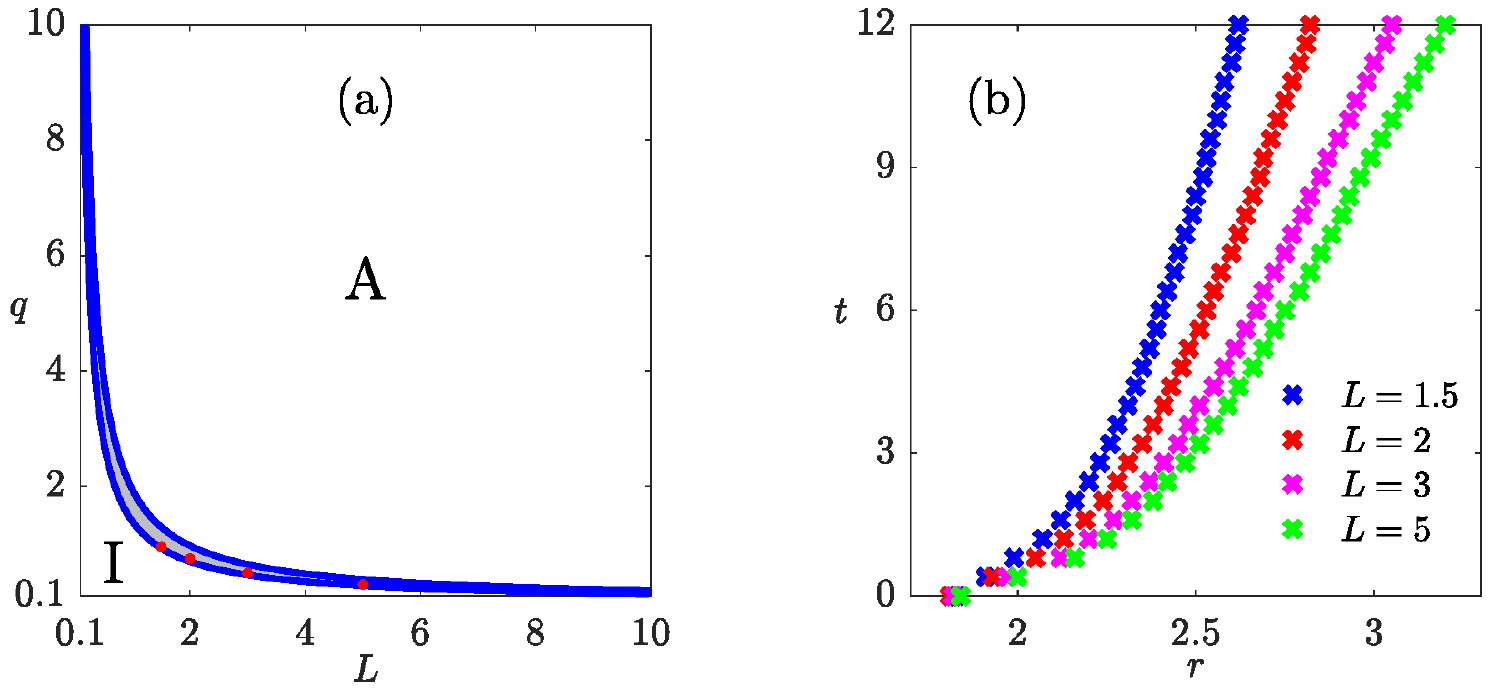}}
	\caption{\footnotesize (a) The upper and lower blue lines mark $\Omega'_{\chi, 0} = 0$ and $\Omega'_{\chi, 0} = 0.15$, respectively. When the wavenumbers $n$ and $k_z$ of a perturbation wave and the swirl number $q$ of the base flow to which the perturbation is added are chosen such that the corresponding case reside in the shaded grey region, the WKB analysis predicts its unbounded, radially outward propagation. On the other hand, when the pair $(L, q)$ falls inside th region labelled by I (A) the WKB analysis signals a continuously inward (a restrained outward) movement. (b) The radius-time plots of cases corresponding to the red points of panel (a), provided by the linear model of Eqs.~(\ref{uh_Fourier}) and (\ref{wh_Fourier}), when $m_f = 2$ and the pair $(L, q)$ is selected as (1.5, 0.95), (2, 0.75), (3, 0.5) and (5, 0.3).} 
	\label{Lq_map}
\end{figure}

\begin{figure}
	\centerline{\includegraphics[width=1.\textwidth]{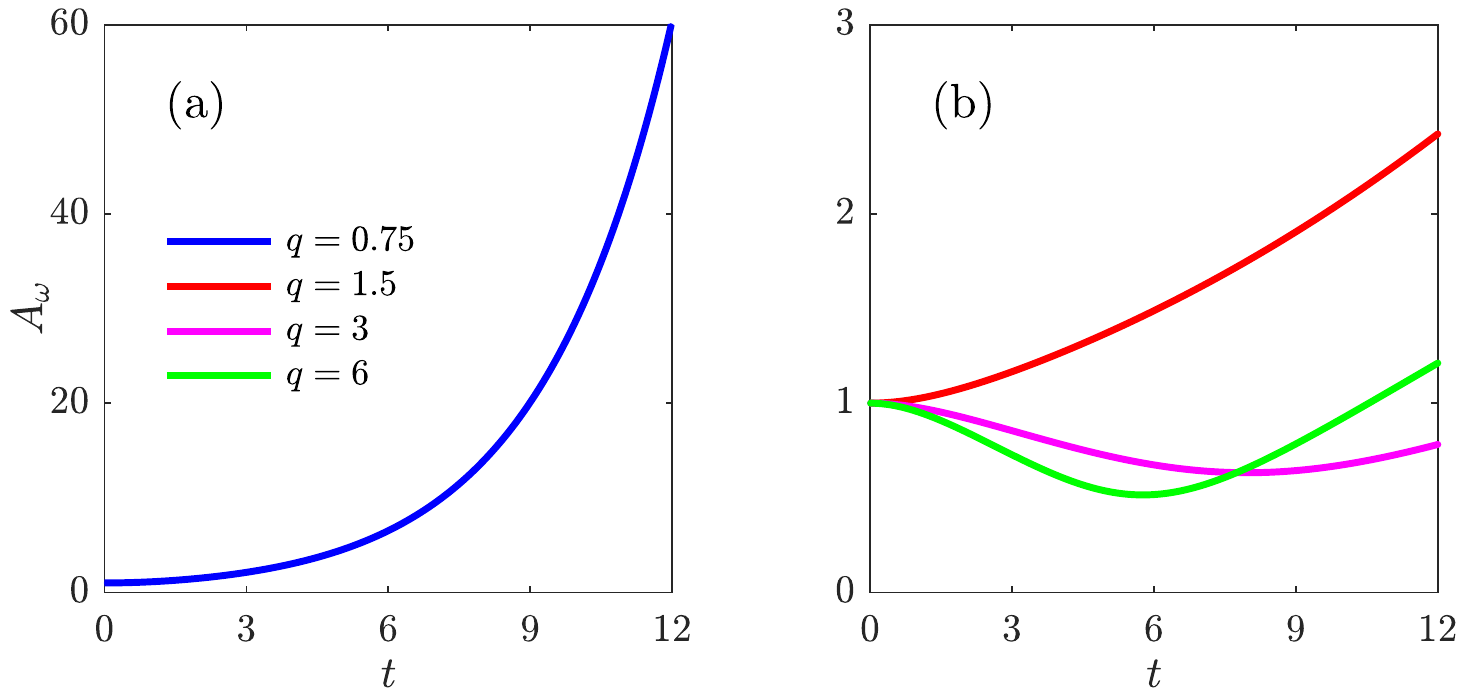}}
	\caption{\footnotesize The WKB predictions for the temporal variation of the amplitude of the helical vorticity perturbation waves for various swirl numbers, when both $m_f$ and $L$ have been maintained constant at 2. For the sake of better visulaization, the results have been shown in two separate panels.} 
	\label{q_effect_amp}
\end{figure}

The WKB predictions largely depend on the value of $\Omega'_{\chi, 0}$, or, strictly speaking, on its sign. When $q$ is finite, $\Omega'_{\chi, 0}$ is formulated as    
\begin{equation}
	\Omega'_{\chi, 0} = \frac{2\exp(-R^2)}{R} - \frac{2\big[1 - \exp(-R^2)\big]}{R^3} + \frac{2R\exp(-R^2)}{qL} \, , \label{wh_imag_simp_allq}
\end{equation}
suggesting that, unless $L \rightarrow 0$ or $L \rightarrow \infty$, there can be found a $q = q_0$ satisfying $\Omega_{\chi, 0}'\big|_{q = q_0} = 0$. Note that, as already discussed, we choose $R = 1$. When $\Omega'_{\chi, 0} > 0$, $k$ steadily decreases with time, leading to $C_{g,1} < 0$ for all $t \geq 0$, thereby causing the continuously inward propagation of perturbation waves and, consequently, $r_1 < 0$ at some finite time, which is physically and mathematically impossible. When $\Omega'_{\chi, 0} \approx 0$ (the shaded grey region in Fig.\ref{Lq_map}a), the radial wavenumber grows very slowly, enabling disturbances to travel much farther from $r = R$, therfore letting the length scale of the wave become comparable to that of the flow field, which can also be viewed as their unbounded outward propagation. On the other hand, when $\Omega'_{\chi, 0} < 0$ and it is noticeably far from zero, the relatively rapid growth of $k$ with time, predicted by the WKB analysis, stagnates the radial group velocity $C_{g1}$ of the helical vorticity waves fairly quickly, leading to the restrained outward movement of the perturbations, as indicated by the asymptotic behaviour of cases with $q \geq 1.5$ in Fig.~\ref{q_effect_rt}(b). Consistent with the findings of the linear model, the WKB analysis also shows that the amplitude of disturbances grows faster with time as $q$ decreases (Fig.~\ref{q_effect_amp}). Furthermore, similar to the linear model, the WKB analysis exhibits that, when $q$ is adequately large (e.g., $q \gtrsim 2$ when $L = 2$), the wave amplitude undergoes some initial decay before reversing the course around the same time that the wave's location reaches its WKB-predicted asymptotic value. We also remark that, although a linear temporal variation has also been established for the radial wavenumber $k(t)$ by the studies employing the WKB approach for the analysis of 3D cyclones in the climate science community \citep{Moller2000, Gao2016}, this variation has been found to be monotonically increasing, since the radial derivative of the basic-state angular velocity at $R$, which is the negative of the rate at which $k(t)$ changes, is always below zero and fairly large in magnitude. This is compatible with the restricted propagation of asymmetries observed in these studies, as the fast and continuous growth of $k$ decelerates the radial and axial group velocities of perturbation waves, and rapidly drives them to zero.     

When $L$ is moderate or large (i.e., $L \gtrsim 1$), the values of $q$ for which the WKB analysis anticipates an unrestrained propagation correspond to low swirl numbers leading to helical instbilities. This suggests that, when $\Omega'_{\chi, 0} \approx 0$ and $L \gtrsim 1$, the linear model of Eqs.~(\ref{uh_Fourier}) and (\ref{wh_Fourier}) should also predict the formation of perturbation waves that travel to large distances from the core radius very rapidly. This can readily be confirmed by comparing the several examples with $\Omega'_{\chi, 0} \approx 0$ presented in Fig.~\ref{Lq_map}(b) with the cases shown in Fig.~\ref{q_effect_rt}(b) for which $q \geq 1.5$. We stress that the shaded area of Fig.~\ref{Lq_map}(a) does not represent the perturbation waves with maximum growth or propagation speed, as further lowering of $q$, for instance, typically increases the amplification rate and the propagation velocity of disturbances. Instead, it merely indicates the region in which the WKB analysis supports the unbounded, radial advection of perturbations, which is a small subset of cases for which the linear model and the numerical simulations predict an unbounded propagation.   

%%%%%%%%%%%%%%%%%%%%%%%%%%%%%%%%%%%%%%%%%%%%%% RDT Analysis %%%%%%%%%%%%%%%%%%%%%%%%%%%%%%%%%%%%%%%%%%%%%%
%%%%%%%%%%%%%%%%%%%%%%%%%%%%%%%%%%%%%%%%%%%%%% RDT Analysis %%%%%%%%%%%%%%%%%%%%%%%%%%%%%%%%%%%%%%%%%%%%%%

\section{Rapid distortion theory \label{section:RDT}}

The linear rapid doistortion theory has proven a powerful tool for the approximation of the velocity and vorticity statistics of a turbulent fields, and for the analysis of vortex-wave interactions. The theory studies the vortex-turbulence interplay by decomposing the flow field to a steady background velocity field $\bs{u}_c(\bs{x})$, which, in this study, is considered to be the 3D Batchelor vortex described in \S \ref{section:Setup}, and an external turbulent velocity field $\bs{u}_e(\bs{x}, t)$, which is initially assumed to be homogenous and isotropic. The turbulent velocity field at $t = 0$ is represented by $\bs{u}^I(\bs{x_0})$. We notice that since fluid elements travel azimuthally and axially in time, the coordinates of $\bs{x}$ and $\bs{x}_0$ are different, so that 
\begin{equation}
	r = r_0  \, , \quad  \theta = \theta_0 + \frac{(1 - \gamma) t}{r^2} \, , \quad z = z_0 + \frac{\gamma t}{q}  \, , \label{RDT_coordinates} 
\end{equation}
where $\gamma = \exp(-r^2)$. The use of RDT is permissible when the two following conditions are fulfilled: 
\begin{enumerate}   
	\vspace{0.025in}
	\item The characteristic velocity of the external flow field defined as $u_s = \sqrt{\big \langle (\bs{u}^I)^2 \big \rangle} (r = r_c)$ is much smaller than that of the columnar vortex, characterized by $\Gamma/r_c$. Here, $\langle \ \rangle$ indicates azimuthal and axial averaging, and $r_c$ and $\Gamma$ denote the radius and circulation of the vortex column, respectively. 
	\item The originally homogenous and isotropic external field becomes progressively inhomogenous and anisotropic, owing to its interactions with the columnar vortex. The resulting inhomogeneous and anisotropic strain rate of the external flow field causes further geometrical distortions on a time scale proprtional to $r_c^2/\Gamma$. A necessary conidition for the implementation of RDT is that $r_c^2/\Gamma \ll t_s = L_s/u_s$, where $L_s$ and $t_s$ are the characteristic length and time scales of the external turbulence.    
\end{enumerate}    

The turbulent velocity field is then divided into a rotational and irrotational (potential) parts such that
\begin{equation}
	\bs{u}_e = \bs{u}^R + \bnabla \phi \, , \label{RDT_decompose}
\end{equation}
where $\phi$ is a velocity potential. The rotational component is related to the initial velocity field by
\begin{equation}
	u_i^R(\bs{x}, t) = \frac{\partial x_{0j}}{\partial x_i} u_{j}^I(\bs{x}_0)  \, , \label{vel_transform_RDT}
\end{equation}
which in the matrix form reads
\begin{equation}
\begin{bmatrix}
u_r^R \vspace{0.05in} \\ 
u_{\theta}^R \vspace{0.05in} \\ 
u_z^R \\
\end{bmatrix} = 
\begin{bmatrix}
1 & a(r,t) & b(r, t) \vspace{0.05in} \\  
0 &   1    &   0     \vspace{0.05in} \\ 
0 &   0    &   1     \\
\end{bmatrix}
\begin{bmatrix}
u_{r}^I \vspace{0.05in} \\
u_{\theta}^I \vspace{0.05in} \\
u_{z}^I \\
\end{bmatrix}   \, . \label{matrix_RDT}
\vspace{0.03in}
\end{equation}
where
\begin{equation}
	a(r, t) = -2\gamma t + \frac{2(1 - \gamma)t}{r^2} \, , \quad b(r, t) = \frac{2r\gamma t}{q}  \, . \label{matrix_coeff}
\end{equation} 
The incompressibility condition also requires
\begin{equation}
	-\nabla^2 \phi = \bnabla \bs{\cdot} \bs{u}^R  \, . \label{continuity_RDT}
\end{equation}
The right handside of Eq.~(\ref{continuity_RDT}) can be linked to the initial velocity field via invoking the chain rule and the matrix relation (\ref{matrix_RDT}) to be evaluated as
\begin{eqnarray}
	\bnabla \bs{\cdot} \bs{u}^R =  a \frac{\partial u_r^I}{\partial \theta_0} + b \frac{\partial u_r^I}{\partial z_0} + \bigg(\frac{a}{r} + \frac{\partial a}{\partial r} \bigg) u_{\theta}^I + a \frac{\partial u_{\theta}^I}{\partial r} + a^2 \frac{\partial u_{\theta}^I}{\partial \theta_0} + \nonumber \\ ab \frac{\partial u_{\theta}^I}{\partial z_0} + \bigg(\frac{b}{r} + \frac{\partial b}{\partial r} \bigg) u_z^I + b \frac{\partial u_z^I}{\partial r} +  ab \frac{\partial u_z^I}{\partial \theta_0} + b^2 \frac{\partial u_z^I}{\partial z_0}  \, . \label{RHS_RDT}
\end{eqnarray}

The Poisson equation of (\ref{continuity_RDT}) is then solved numerically along with the boundary conditions 
 \begin{eqnarray}
	\bs{u}_e &\rightarrow& 0 \quad \mrm{as} \ r \rightarrow \infty \, , \nonumber \\ 
	\bs{u}_e &=& 0 \quad \mrm{at} \ r = 0 \, , \label{BC_RDT}
\end{eqnarray}
using the successive over-relaxation method on a computational grid stretching over $L_x \times L_y \times L_z = 20^3$ with the mesh size $512^3$. Second-order centered schemes have been used to discretize the derivatives of Eq.~(\ref{continuity_RDT}). The initial isotropic  field $\bs{u}^I$ is generated by adopting the approach proposed in \citet{Kwak1975}. The reynolds stresses ${u}_r^{\prime 2}$ and $u_r' u'_{\theta}$, whose DNS-based counterparts will be presented in the next section, are then simply calculated as $u_{e,r}^2$ and $u_{e,r} u_{e, \theta}$, respectively. 

\begin{figure}
	\centerline{\includegraphics[width=1.\textwidth]{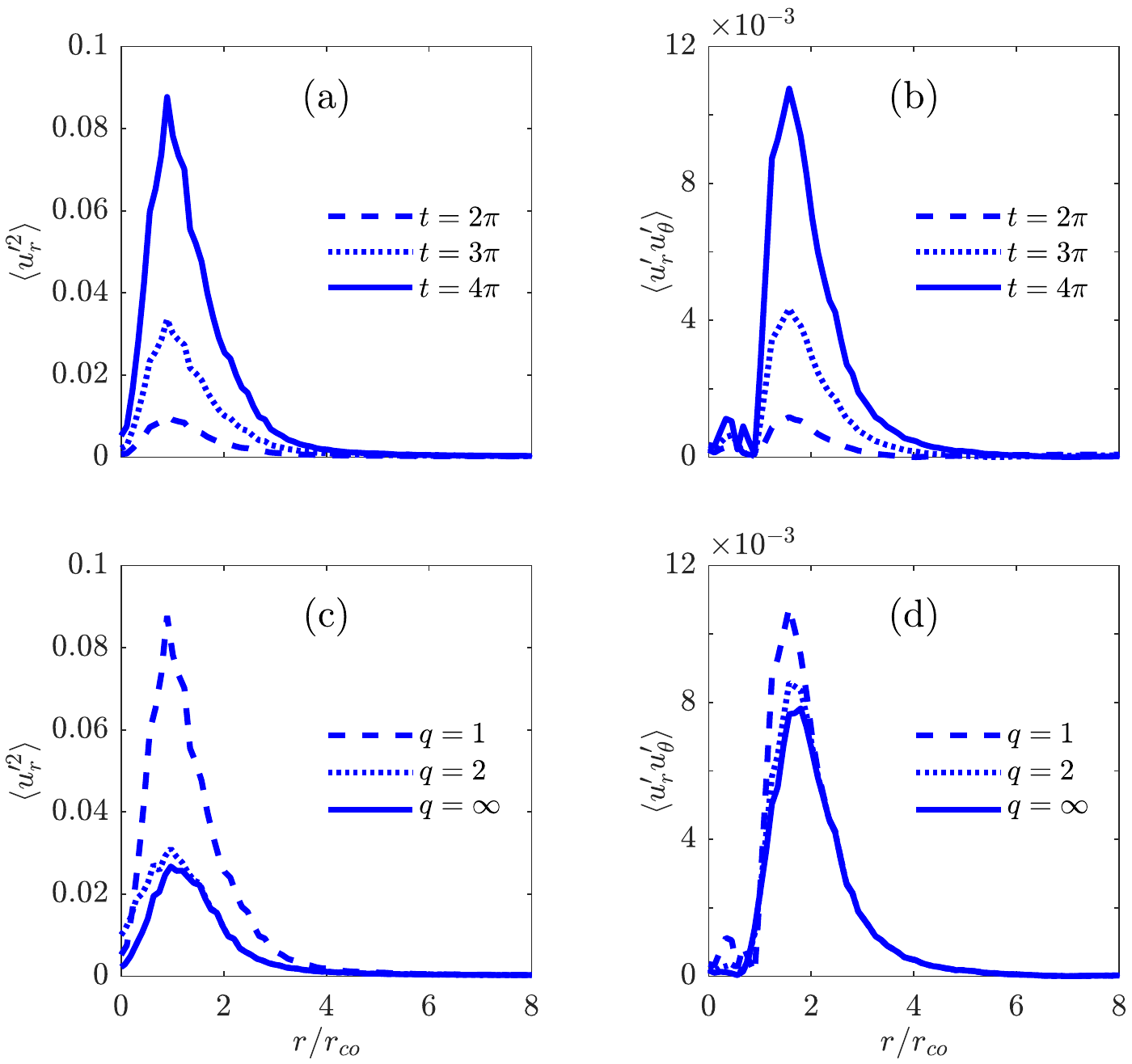}}
	\caption{\footnotesize Top: Azimuthally and axially averaged Reynolds stresses (a) $\langle u_r^{\prime 2} \rangle$, and (b) $\langle u_r' u'_{\theta} \rangle$, evaluated at several different times for a flow with $q = 1$. Bottom: Same as the first row, but calculated for various swirl numbers at the constant time $t = 4\pi$.} 
	\label{RDT}
\end{figure}

The dominance of $\big \langle {u}_r^{\prime 2} \big \rangle$ at small radii, shown in Fig.~\ref{RDT}(a) and (c), suggests that, consistent with the findings of previous studies (e.g., \citealp{Miyazaki2000, Duraisamy2008}), momentum transport in the vicinity of the vortex core is mainly governed by Reynolds normal stresses, through which disturbances are carried radially outward in this region of the flow. The RDT analysis further exhibits that $\big \langle {u}_r^{\prime 2} \big \rangle$ peaks inside the core (at $0.9 \lesssim r/r_{co}\lesssim 0.95$ depending on the swirl number), and decays rapidly to zero outside it. It can also be observed that the peak's location barely varies with time, and only marginally shifts to larger radii as $q$ increases (Figs.~\ref{RDT}a and c). It can be further noted that the maximum value of $\big \langle {u}_r^{\prime 2} \big \rangle$ grows with time and dwindles with swirl number, while the latter effect corresponds to the attenuation of modal instabilities. On the other hand, the steadily growing peak of $\langle u_r' u'_{\theta} \rangle$ is located noticeably outside the core (Fig.~\ref{RDT}b and d), revealing that, at large radii, the transport of angular momentum is largely predominated by the generation of $\langle u_r' u'_{\theta} \rangle$, which in turn facilitates the outward propagation of perturbation waves. Similar to $\big \langle {u}_r^{\prime 2} \big \rangle$, the maximum value of $\langle u_r' u'_{\theta} \rangle$ increases with the decrease of swirl number (Fig.~\ref{RDT}d). Furthermore, RDT indicates that the peak's location is again insensitive to time, but travels slightly away from the core as the swirl number grows. For instance, the radial location of the maximum changes from $r/r_{co} \approx 1.55$ to $r/r_{co} \approx 1.75$, as $q$ varies increases from $1$ to $\infty$ (Fig.~\ref{RDT}d). 

%%%%%%%%%%%%%%%%%%%%%%%%%%%%%%%%%%%%%%%%%%%%%% DNS Results %%%%%%%%%%%%%%%%%%%%%%%%%%%%%%%%%%%%%%%%%%%%%%
%%%%%%%%%%%%%%%%%%%%%%%%%%%%%%%%%%%%%%%%%%%%%% DNS Results %%%%%%%%%%%%%%%%%%%%%%%%%%%%%%%%%%%%%%%%%%%%%%

\section{DNS results \label{section:DNS}}

In this section, the DNS solver described in \S \ref{section:Setup} is utilized to elucidate the underlying physics of the outward advection of perturbations, and to determine how the findings of the RDT analysis presented earlier compare with those of a more intricate and comprehensive numerical model. As discussed in \S \ref{section:Setup}, the numerical solver is set up for both the full and linearized governing equations, while the latter is called L-DNS, and is shown to be intimately tied to the linear RDT analysis of the previous section \citep{Pradeep2006}. L-DNS enables the study of transient growth mechanisms at a given finite swirl number, since it retains the swirl number at its prescribed value at $t = 0$. On the other hand, when the full DNS of the flow is performed, $q$ tends to rapidly increase with time, unless it is initially set at a very large number (infinity).  

\begin{figure}
	\centerline{\includegraphics[width=1.\textwidth]{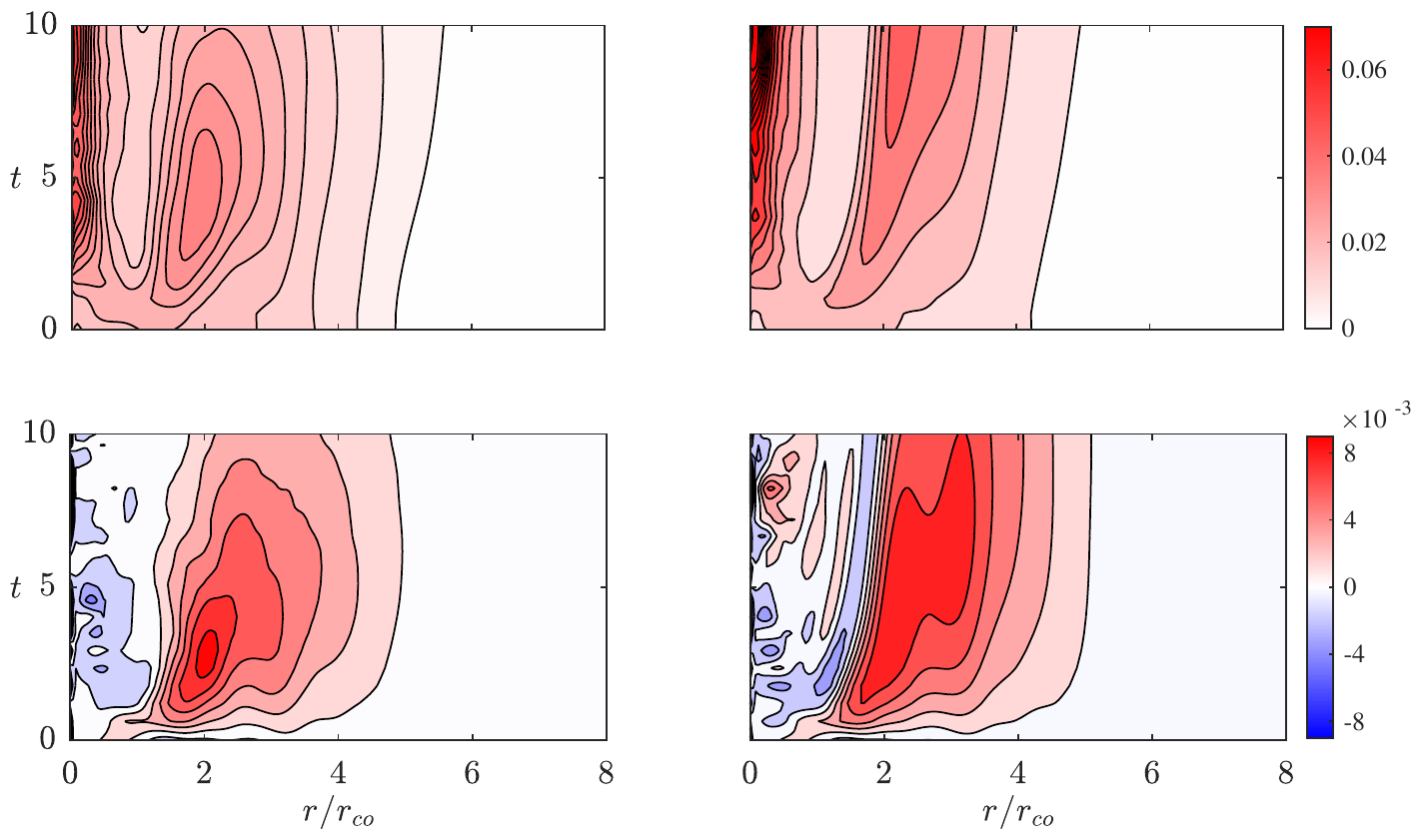}}
	\caption{\footnotesize The spatiotemporal evolution of the azimuthally and axially averaged Reynolds stresses $\langle u_r^{\prime 2} \rangle$ (top) and $\langle u_r' u'_{\theta} \rangle$ (bottom), provided by the full DNS of the flow (right) and the DNS of its linearized transport equations, when $q = \infty$.} 
	\label{DNS_vs_LNS}
\end{figure}

Before leveraging L-DNS to uncover the primary mechanisms resposible for the outward propagation of asymmetries, it is of paramount importance to first establish its validity by comparing its predictions with those of the full DNS. The spatiotemporal evolution of the radially and azimuthally averaged Reynold stresses $\langle u_r^{\prime 2} \rangle$ (top) and $\langle u_r' u'_{\theta} \rangle$ (bottom) provided by DNS (left) and L-DNS (right) are depicted in Fig.~\ref{DNS_vs_LNS}, when $q = \infty$. This is the only choice of swirl number for which a fair comparison between L-DNS and the full DNS is possible. The qualitative agreement between the two in terms of revealing the underlying mechanisms of momentum transport is striking. Both models show that $\langle u_r^{\prime 2} \rangle$ is mainly contained to the vicinity of the core, while indicating the advection of $\langle u_r' u'_{\theta} \rangle$ to larger radii with time. The global maximum of $\langle u_r' u'_{\theta} \rangle$ is found to be located at $r/r_{co} \approx 2$ and $t \approx 3$ by both the full DNS and L-DNS, also manifesting their reasonable quantitative agreement. All these observations are generally consistent with the results of the RDT analysis of \S \ref{section:RDT}. The major discrepancy between the two numerical models lies in the faster rate of decay predicted by the full DNS for $u_r^{\prime 2}$ and $u_r' u'_{\theta}$ (as well as vorticity perturbations) owing to nonlinear processes that are naturally ignored by a linear model such as L-DNS. It should be noted that, due to the limitations imposed by the mesh size, the results at small values of $r$ (near the origin) are discernibly noise-corrupted, and thus should be viewed and interpreted cautiously.   

\begin{figure}
	\centerline{\includegraphics[width=1.\textwidth]{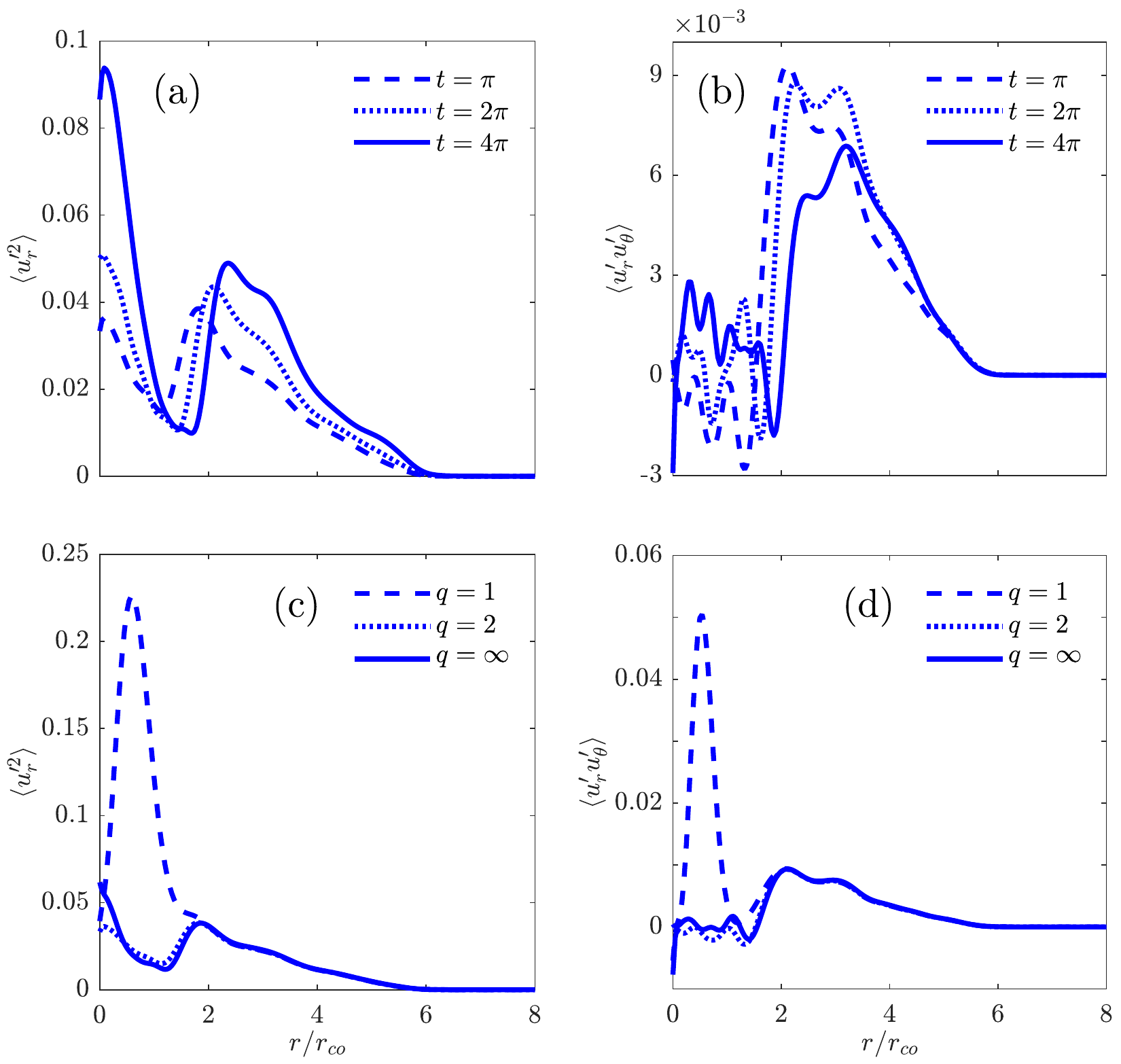}}
	\caption{\footnotesize Top: Reynolds stresses (a) $\langle u_r^{\prime 2} \rangle$, and (b) $\langle u_r' u'_{\theta} \rangle$, obtained from the L-DNS of a flow whose swirl number is set at $q = 2$, and displayed at three distinct times. Bottom: Same as top, but for different values of $q$, while time is fixed at $t = \pi$.} 
	\label{DNS_RS}
\end{figure}

\begin{figure}
	\centerline{\includegraphics[width=1.\textwidth]{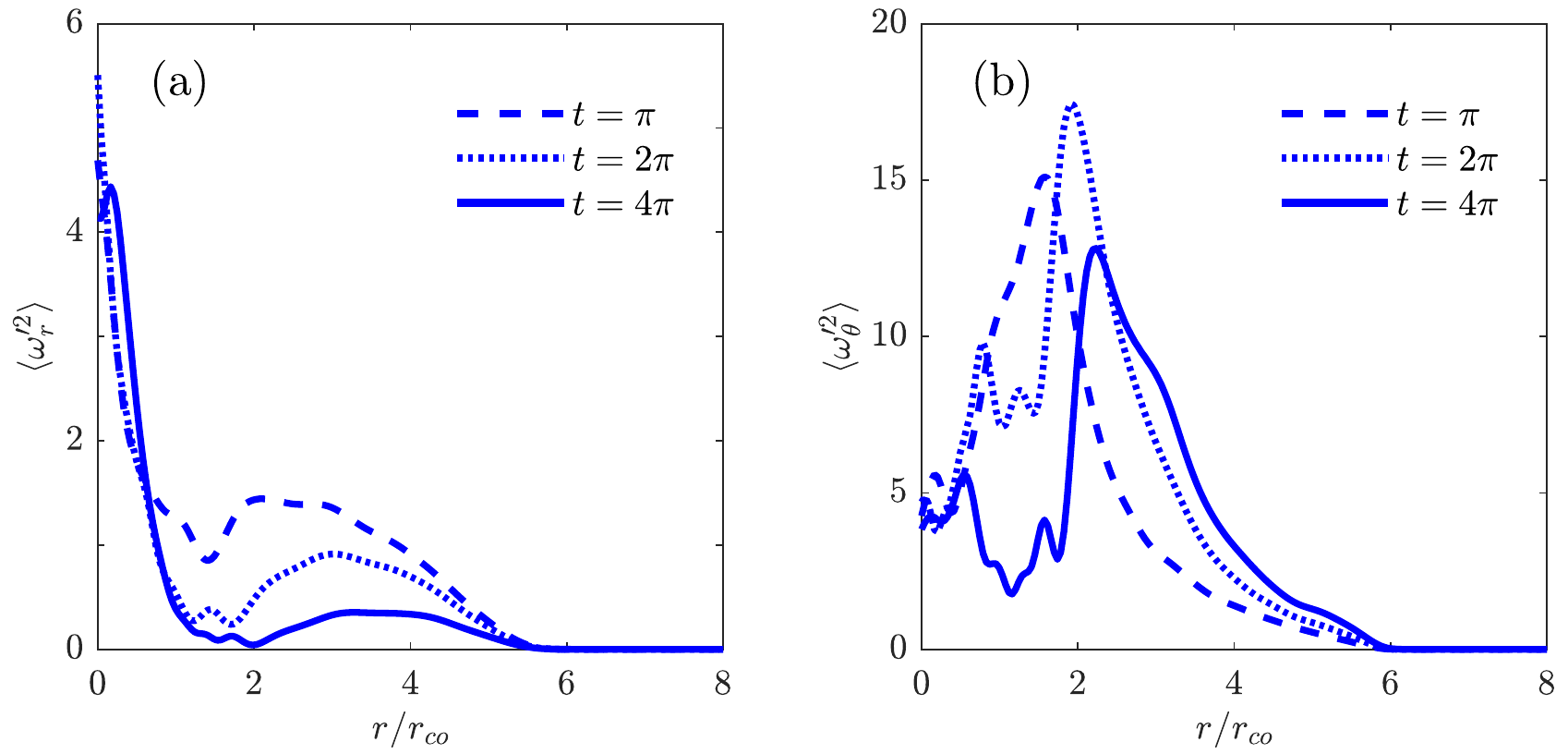}}
	\caption{\footnotesize The radial variation of (a) $\langle \omega_r^{\prime 2} \rangle$, and (b) $\langle \omega_{\theta}^{\prime 2} \rangle$ at several times, rendered by the L-DNS of a flow with $q = 2$.} 
	\label{DNS_vort}
\end{figure}

Overall, as can be seen in Figs.~\ref{DNS_RS}(c) and (d), L-DNS demonstrates a very similar behaviour for all flows with $q \geq 1.5$. Again, in accordance with the RDT-based results, L-DNS shows that $u_r^{\prime 2}$ is typically maximized inside the core, at least for large enough times (i.e., $t \gtrsim 1.5 \pi$), and for $u_r' u'_{\theta}$ to peak outside it, while, in contrast with the RDT-based findings of the previous section, its peak constantly shifts toward larger radii with time (Figs.~\ref{DNS_RS}a and b). The L-DNS results also indicate the existence of a local maximum for $\langle u_r^{\prime 2} \rangle$ outside the core, which is smaller than the peak appearing within the core if $t \gtrsim 1.5 \pi$, which is another qualitative difference between the predictions of L-DNS and the RDT analysis of \S \ref{section:RDT}. The tilting and stretching mechanism of \citet{Pradeep2006}, proposed for the transient growth of perturbations, is corroborated by Fig~\ref{DNS_vort}, as it illustrates the constant reduction of radial vorticity with time due to the tilting of radially aligned vortex lines toward azimuthal direction, and the growth of azimuthal vorticity caused by the stretching of radial vorticity by the background strain. As shown in this figure, the further depletion of $\omega_r$, initially decelerates and then reverses the growth of $\omega_{\theta}$, which in turn terminates the production of $u_r' u'_{\theta}$, leading to the arrest of the transient amplification of perturbations and their outward propagation. Here, $\langle \omega_r^{\prime 2} \rangle$ and $\langle \omega_{\theta}^{\prime 2} \rangle$ are chosen to assess the decay and growth of various vorticity components, since they are notably less affected by noise than than $\langle \omega_r' \rangle$ and $\langle \omega_{\theta}' \rangle$, and therefore are capable of providing a more reliable picture of the physical processes in action. 

\begin{figure}
	\centerline{\includegraphics[width=1.\textwidth]{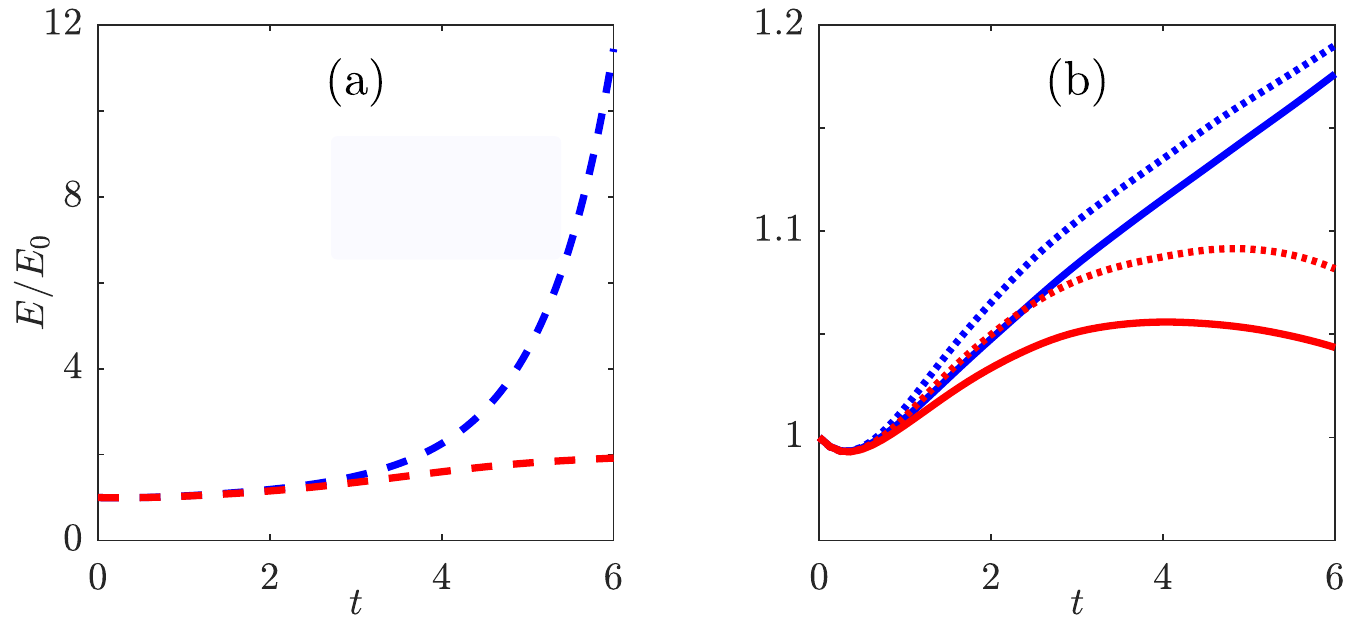}}
	\caption{\footnotesize The temporal growth of the turbulent kinetic energy $E$, scaled by its initial value $E_0$, for $q = 1$ (dashed), $q = 2$ (dotted) and $q = \infty$ (solid). The blue and red lines represent the results of L-DNS and N-DNS, respectively. For the sake of a clearer distinction between the findings at different swirl numbers, the results for the case with modal growth ($q = 1$) has been separated from the other two with only algebraic growth.} 
	\label{TKE}
\end{figure}

The growth mechanisms described above are enhanced for flows with $q \lesssim 1.5$ (dashed lines in Fig.~\ref{DNS_RS}c and d), due to the presence of helical instabilities, resulting in the modal exponential growth of disturbances (in the absence of nonlinear interactions), and therefore their faster radial propagation, as discussed in earlier sections. This can be further highlighted by examining the temporal variation of turbulent kinetic energy (TKE) $E$ calculated as   
\begin{equation}
	E(t) = \frac{1}{L_xL_yL_z} \int_0^{L_z} \int_0^{L_y} \int_0^{L_x} \frac{1}{2}\big(u^{\prime 2} + v^{\prime 2} + w^{\prime 2}\big) \ \mrm{d}x \ \mrm{d}y \ \mrm{d}z \, ,\label{TKE_eqn}   
\end{equation}     
where $u'$, $v'$ and $w'$ are velocity perturbations in $x$-, $y$- and $z$-directions of Cartesian coordinates, respectively. $E_0$, used for the normalization of TKE, is $E$ at $t = 0$. While the anticipated exponential amplification of TKE for the linear flow with $q = 1$ is confirmed in Fig.~\ref{TKE}(a), the linear, normal-mode-stable cases, indicated by blue lines in Fig.~\ref{TKE}(b), solely exhibit an algebraic energy growth, because of the transient effects that vanish at large times. 

\subsection{Nonlinear effects and growth arrest \label{section:NL}}

The present RDT- and L-DNS-based analyses provide clear-cut phenomenological explanations for the mechanisms enabling the radial advection of disturbances in 3D vortex-dominated flows, which was theoretically quantified by the linear model of \S \ref{section:linear_model} and its associated WKB-based reduction. Not surprisingly, however, both RDT and L-DNS fall short to shed light on the nonlinear mechanisms accounting for the cessation of the growth and outward propgation of perturbations, as they only capture the linear turbulence-mean-flow interactions. The N-DNS numerical approach, discussed in \S \ref{section:Setup}, however, can be leveraged to study the nonlinear phenomena neglected by RDT and L-DNS, as unlike the two, N-DNS integrates turbulence-turbulence interactions, while freezing $q$ at its initial value by discarding the base-flow interactions, and hence the mean flow changes.  

\begin{figure}
	\centerline{\includegraphics[width=1.\textwidth]{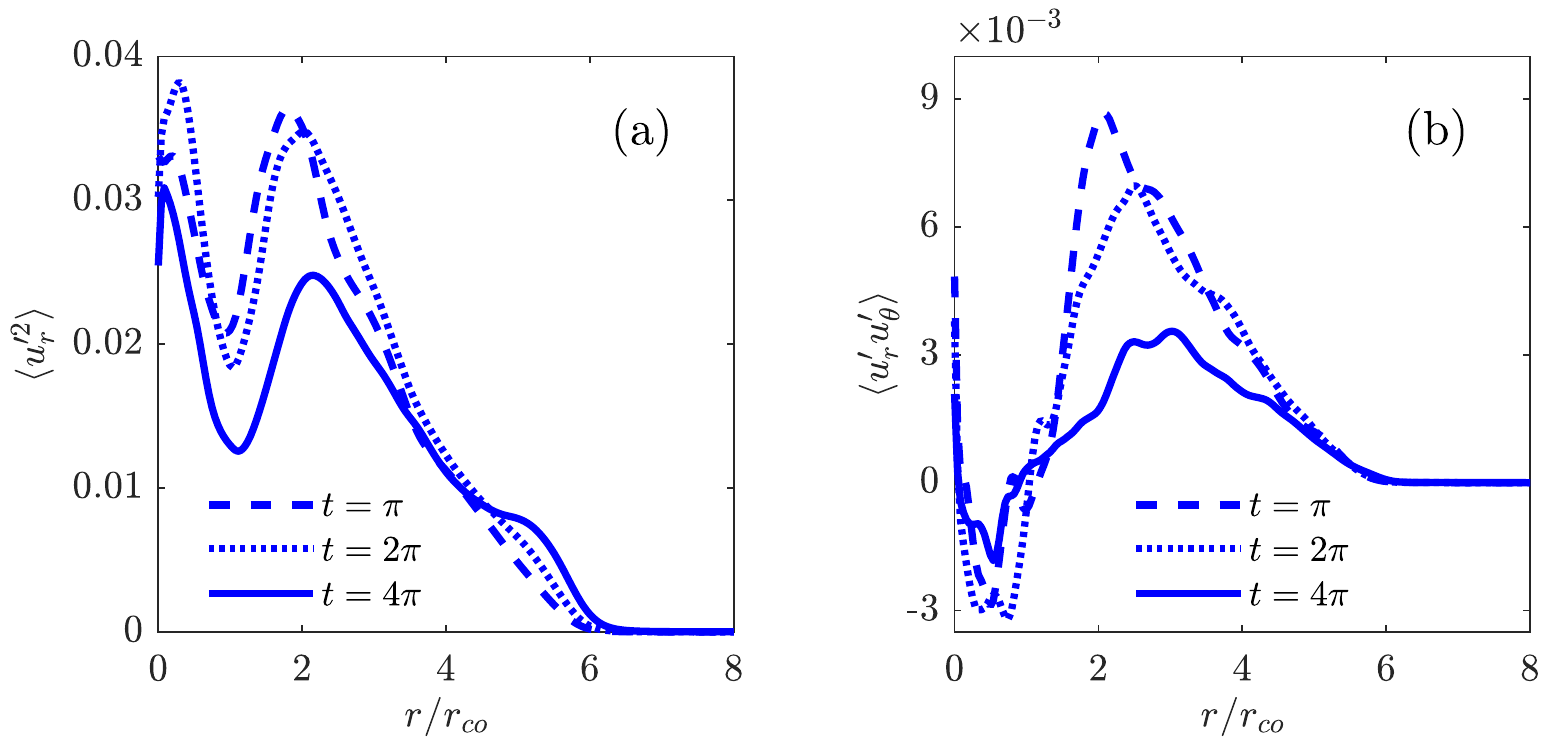}}
	\caption{\footnotesize The radial distribution of the Reynolds stresses $\langle u_r^{\prime 2} \rangle$ (left) and $\langle u_r' u'_{\theta} \rangle$ (right) at different times, given by the N-DNS of a perturbed Batchelor vortex, when the swirl number is taken as $q = 2$.} 
	\label{NL_RS}
\end{figure}   

The radial and temporal variations of the Reynolds stresses $\langle u_r^{\prime 2} \rangle$ (Fig.~\ref{NL_RS}a) and $\langle u_r' u'_{\theta} \rangle$ (Fig.~\ref{NL_RS}b) rendered by the N-DNS of the flow with $q = 2$ can be contrasted against the correponding L-DNS case presented in Figs.~\ref{DNS_RS}(a) and (b). While similar to the linear simulations, the out-of-the-core maximum of  $\langle u_r^{\prime 2} \rangle$ and the peak of $\langle u_r' u'_{\theta} \rangle$ gradually move to larger radii, both $\langle u_r^{\prime 2} \rangle$ and $\langle u_r' u'_{\theta} \rangle$ diminish at much faster rates in comparison with the corresponding L-DNS results. In fact, for the considered time span of $t = \pi$ to $4\pi$, L-DNS does not reveal any decline for the Reynolds normal stress  $\langle u_r^{\prime 2} \rangle$. These observations are fully compatible with the mechanism proposed for the disruption of energy growth in Lamb-Oseen vortices (Batchelor vortices at $q = \infty$) by \citet{Hussain2011}, wherein the reduction of $\omega'_{\theta}$, and hence $\langle u_r' u'_{\theta} \rangle$, is accelereted by the nonlinear interaction between the vortex cells as well as the induced velocity of the vortex filaments that roll the cells up and drive them away from the vicinity of the core, thereby removing disturbances from the neighborhoud (resonant radius) around which the optimal growth modes are localized. The substantial contribution of nonlinear effects to the arrest of transient growth is further substantiated by the energy plots of Fig.~\ref{NL_RS}, displaying the slower amplification and the eventual decay of TKE for the nonlinear flows (red lines), including when $q < 1.5$, compared with the continuous exponential (if $q < 1.5$) or algebraic growth of TKE in the linear flows (blue lines).  

%%%%%%%%%%%%%%%%%%%%%%%%%%%%%%%%%%%%%%%%%%%%%% Conclusions %%%%%%%%%%%%%%%%%%%%%%%%%%%%%%%%%%%%%%%%%%%%%%
%%%%%%%%%%%%%%%%%%%%%%%%%%%%%%%%%%%%%%%%%%%%%% Conclusions %%%%%%%%%%%%%%%%%%%%%%%%%%%%%%%%%%%%%%%%%%%%%%

\section{Conclusions \label{section:Conclusions}}

 The spatiotemporal evolution of asymmetries introduced to 3D turbulent trailing vortices was studied using a variety of analytical and numerical methodologies. A 2D inviscid model was developed by invoking the helical symmetry of these flows and linearizing their transport equations for momentum and vorticity around the base flow, which was taken to be a Batchelor vortex. The PDEs of this linear model were further simplified to a handful of ODEs using the WKB analysis, which relies on approximating disturbances as tightly-wound wavepackets. While the linear model indicated that asymmetries added to 3D vortices can travel unboundedly outward for all values of $q$, in contrast with the behaviour observed for 2D flows, the WKB approach, as a normal-mode stability analysis technique, unsurprisingly neglected the non-modal mechanisms promoting transient growth and the radially outward advection of perturbations except for a narrow range of swirl numbers and perturbation wavenumbers (Fig.~\ref{Lq_map}a), which is also consistent with the compactness assumption made by this method for perturbation waves.
 
 In addition, we investigated the physical processes responsible for the outward propagation of perturbations using RDT- and DNS-based analyses. When $q \gtrsim 1.5$, the same inviscid mechanisms as those proposed by \citet{Pradeep2006} and \citet{Heaton2007} for transient growth are found to be at play, where the tilting of $\omega_r$ into $\omega_{\theta}$ and the subsequent streatching and increase of $\omega_{\theta}$ by the mean strain field generate positive $u_r' u'_{\theta}$, which in turn facilitates the radially outward propagation of disturbances. In line with the findings of the 2D linear model and the WKB analysis, for $q \lesssim 1.5$, the presence of helical instabilities enables the modal (exponential) growth for the amplitude of disturbances and the kinetic energy of the system, resulting in the emergence of the rapidly-growing and -travelling perturbation waves.   
 
 The RDT methodology and the L-DNS numerical model adopted in this study gave valuable insights into the underpinnings of momentum transport and the radial advection of disturbances at early stages of the flow, but due to their linear nature, they were unable to describe the nonlinear mechanisms leading to vortex breakdown, the disruption of transient growth and the arrest of energy amplification occurring at later times. N-DNS, on the other hand, enabled the exploration of the nonlinear physics at finite swirl numbers by including turbulence-turbulence interations and retaining $q$ at its initial value. The comparison between the results of L-DNS and N-DNS illustrated that the nonlinear effects expedite the reduction of $u_r'u_{\theta}'$, as they contribute to the radially outward self-advection of vortex dioples and, as a result, to the migration of perturbations away from the core, where the weaker mean strain rate cannot sustain their growth and outward propagation. Although to a lesser extent than the inviscid nonlinear interactions, the viscous dissipation can also slow down the growth rates and the outward advection of asymmetries. The effect of $Re$ on the radial propagation of perturbations will be studied in detail in a forthcoming work through numerical simulations and the introduction of viscous diffusion to the linear model of \S \ref{section:Theory} and its concomitant WKB analysis. 

%%%%%%%%%%%%%%%%%%%%%%%%%%%%%%%%%%%%%%%%%%%%%% Acknowledgment %%%%%%%%%%%%%%%%%%%%%%%%%%%%%%%%%%%%%%%%%%%%%%
%%%%%%%%%%%%%%%%%%%%%%%%%%%%%%%%%%%%%%%%%%%%%% Acknowledgment %%%%%%%%%%%%%%%%%%%%%%%%%%%%%%%%%%%%%%%%%%%%%%

\section*{Acknowledgment \label{section:Acknowledgment}}

M. A. Khodkar thanks Karthik Duraisamy and Pedram Hassanzadeh for fruitful discussions, insightful comments and sharing their computational resources.

\section*{Declaration of interestst \label{section:Conflict_of_Interest}}

The author reports no conflict of interest.

% The reference section via BibTeX
\bibliographystyle{jfm}
\bibliography{Main}

\end{document}